\documentclass{aa}  
\newcommand{\appropto}{\mathrel{\vcenter{
  \offinterlineskip\halign{\hfil$##$\cr
    \propto\cr\noalign{\kern2pt}\sim\cr\noalign{\kern-2pt}}}}}
\usepackage{graphicx}
\usepackage[varg]{txfonts}
\usepackage[colorlinks=true, citecolor=blue]{hyperref}
\DeclareFontFamily{U}{wncy}{}
\DeclareFontShape{U}{wncy}{m}{n}{<->wncyr10}{}
\DeclareSymbolFont{mcy}{U}{wncy}{m}{n}
\DeclareMathSymbol{\Sh}{\mathord}{mcy}{"58} 

\usepackage[framemethod=tikz]{mdframed}
\definecolor{mycolor}{rgb}{0.122, 0.435, 0.698}
\newmdenv[innerlinewidth=0.5pt, roundcorner=4pt,linecolor=mycolor,innerleftmargin=6pt,
innerrightmargin=6pt,innertopmargin=6pt,innerbottommargin=6pt]{mybox}

\begin{document}

   \title{Trade-offs in high-contrast integral field spectroscopy for exoplanet detection and characterisation}

   \subtitle{Young gas giants in emission}
   \author{R. Landman\inst{1}, I.A.G. Snellen\inst{1}, C.U. Keller\inst{1,2}, M. N'Diaye\inst{3}, F. Fagginger-Auer\inst{1}, C. Desgrange\inst{4,5}}
    \authorrunning{R. Landman et al.}

   \institute{Leiden Observatory, Leiden University, Postbus 9513, 2300 RA Leiden, The Netherlands \\
   \email{rlandman@strw.leidenuniv.nl}
   \and
   Lowell Observatory, 1400 W Mars Hill Rd Flagstaff, AZ 86001, USA
   \and
   Université Côte d’Azur, Observatoire de la Côte d’Azur, CNRS, Laboratoire Lagrange, 06108 Nice, France
   \and
 Universit\'e Grenoble Alpes, CNRS, IPAG, 38000 Grenoble, France
   \and 
    Max Planck Institute for Astronomy, K\"onigstuhl 17, D-69117 Heidelberg, Germany}

   \date{Received XXX; accepted YYY}

 
  \abstract
    {Combining high-contrast imaging with medium- or high-resolution integral field spectroscopy has the potential to boost the detection rate of exoplanets, especially at small angular separations. Furthermore, it immediately provides a spectrum of the planet that can be used to characterise its atmosphere. The achievable spectral resolution, wavelength coverage, and FOV of such an instrument are limited by the number of available detector pixels.}
   {We aim to study the effect of the spectral resolution, wavelength coverage, and FOV on the detection and characterisation potential of medium- to high-resolution integral field spectrographs with molecule mapping.}
   {The trade-offs are studied through end-to-end simulations of a typical high-contrast imaging instrument, analytical considerations, and atmospheric retrievals. The results are then validated with archival VLT/SINFONI data of the planet $\beta$ Pictoris b.}
   {We show that molecular absorption spectra generally have decreasing power towards higher spectral resolution and that molecule mapping is already powerful for moderate resolutions (R$\gtrsim$300). When choosing between wavelength coverage and spectral resolution for a given number of spectral bins, it is best to first increase the spectral resolution until R$\sim$2,000 and then maximise the bandwidth within an observing band. We find that T-type companions are most easily detected in the J/H band through methane and water features, while L-type companions are best observed in the H/K band through water and CO features. Such an instrument does not need to have a large FOV, as most of the gain in contrast is obtained in the speckle-limited regime close to the star. We show that the same conclusions are valid for the constraints on atmospheric parameters such as the C/O ratio, metallicity, surface gravity, and temperature, while higher spectral resolution (R$\gtrsim$10,000) is required to constrain the radial velocity and spin of the planet.}
   {}


    \maketitle

%

\section{Introduction}
Direct imaging of exoplanets provides an ideal way to study their atmospheres and orbital configurations \citep{2016PASP_bowler_direct_imaging}. These observations are technically challenging, as they require overcoming a large contrast between a bright star and a dim exoplanet at a small angular separation. This can be achieved through a combination of extreme adaptive optics (XAO), coronagraphy, and post-processing. Imaging surveys with dedicated high-contrast imaging instruments on 8-metre class telescopes, such as SPHERE \citep{2019A&A_beuzit_sphere}, GPI \citep{2014PNAS_macintosh_gpi} and SCExAO \citep{2015PASP_jovanovic_scexao}, have resulted in the detection of a few dozen planetary and brown dwarf companions \citep[e.g.][]{2015Sci_macintosh_51eri,2016ApJ_konopacky_hr2562_substellar_companion, 2017A&A_chauvin_hip65426, 2017A&A_milli_hd206893, 2018A&A_keppler_pds70,  2018A&A_cheetham_hip64892_brown_dwarf,2020ApJ_Currie_brown_dwarf, 2022NatAs_Currie_AB_Aur, 2020MNRAS_Bohn_yses1, 2021A&A_bohn_yses2}. These surveys have mostly been sensitive to young, wide-orbit planets that are still luminous from their formation. While the surveys provide constraints on the demographics of these widely separated planets \citep{2019AJ_nielsen_gpies,2021A&A_vigan_shine}, these companions appear to be relatively rare. To probe the bulk of the giant planet population, we need to increase the contrast, especially at smaller angular separations. However, most of the currently used post-processing techniques, such as angular differential imaging (ADI) and spectral differential imaging (SDI) perform poorly at small angular separations, < 0.2" \citep{2006ApJ_marois_adi, 2015A&A_rameau_sdi}. These algorithms fundamentally suffer from this limitation because the amount of diversity decreases with a smaller angular separation.

In contrast, high-dispersion spectroscopy can detect thermal emission from both transiting and non-transiting exoplanets with minimal spatial separation and starlight suppression by cross-correlating with planet-model spectra \citep[e.g.][]{2012Natur_brogi_tau_bootis, 2014Natur_snellen_beta_pic_crires}. This technique can use both the rapidly varying radial velocity of the planet as well as the distinct planetary spectral features to distinguish between the light of the planet and the star. High-dispersion spectroscopy has been used to detect molecules in exoplanet atmospheres \citep[e.g.][]{2013MNRAS_birkby_water_hd189, 2013A&A_deKok_CO_HD189, 2014A&A_brogi_hd179_co_h2o} and, for example, to estimate their spin \citep{2014Natur_snellen_beta_pic_crires,2016A&A_schwarz_spin_gq_lup, 2018NatAs_Bryan_spin, Wang_2021AJ_hr8799_kpic}. \citet{2002ApJ_sparks_ford_mm} were the first to propose that the combination of high-contrast imaging with a medium to high spectral resolution integral field spectrograph (IFS) can improve our ability to detect exoplanets. The potential of this technique has been studied many times \citep{2015A&A_snellen_hdc_hci, 2007A&A_riaud_hci+hdc_eartlike, 2017AJ_wang_high_dispersion_coronagraphy}. While current high-contrast imagers only have low spectral resolution IFSs (R<100), observations using instruments such as VLT/SINFONI, Keck/OSIRIS, and VLT/MUSE (R>1,000) have indeed demonstrated the potential of these techniques \citep{2013Sci_konopacky_hr8799_osiris, 2018A&A_hoeijmakers_mm, 2019NatAs_haffer_pds70}, even though those instruments were not designed for high-contrast imaging. Not only can such IFSs be used to boost the detection limits, but they can also simultaneously provide a moderate-resolution spectrum of the planet, thereby facilitating both detection and atmospheric characterisation with a single observation. Moderate-resolution emission spectra obtained using these instruments have been used to detect molecules \citep[e.g.][]{2013Sci_konopacky_hr8799_osiris, 2015ApJ_barman_hr8799_osiris, 2018A&A_petit_hr8799_osiris, 2018A&A_hoeijmakers_mm}, accretion tracers \citep{2019NatAs_haffer_pds70, 2020A&A_Eriksson_Halpha}, and even isotopes \citep{2021Natur_zhang_isotope}. Furthermore, they can  constrain the radial velocity \citep{2019AJ_ruffio_radial_velocity} and atmospheric parameters, such as the C/O ratio and metallicity \citep[e.g.][]{Ruffio_2021AJ_HR8799_deep, 2021A&A_petrus_hip65426b, 2020AJ_wilcomb_kappa_and_b_osiris}, thereby revealing information about the formation and migration history of the planet \citep{Molliere2022arXiv_formation_composition}.

These promising results have led to the development of multiple instruments that aim to combine high-contrast imaging and high-dispersion spectroscopy. These include the coupling of existing high-contrast imagers with high-dispersion spectrographs such as VLT/HIRISE \citep{2018SPIE10702E_vigan_hirise,2021A&A_otten_hirise}, Keck/KPIC \citep{2019arXiv_jovanovic_kpic, 2021JATIS_delorme_kpic}, and Subaru/REACH \citep{2017arXiv_jovanovic_reach, 2020SPIE11_kotani_subaru_reach}. However, these instruments only use a single or a few fibers in the focal plane. If the position of the planet is known, they allow for detailed characterisation of the planets, as was done for the HR8799 planets in \citet{Wang_2021AJ_hr8799_kpic}. While new techniques are being developed to use such instruments for planet detection within the diffraction limit \citep{2020SPIE11446E..19E_Echeverri_VFN, 2022ApJ...938..140X_Xin_photonic_lantern_nulling}, these instruments generally do not allow for the search of unknown companions over a significant FOV. In such cases, a larger FOV needs to be sampled through an IFS. Promising upcoming instruments with such a capability are ERIS \citep{2018SPIE10702E_davies_eris} and MedRes, the medium-resolution IFS for the planned upgrade of SPHERE \citep{2020arXiv_boccaletti_sphere+, 2022Spie_Gratton_MedRes}. The arrival of the ELT, HARMONI \citep{2021Msngr_Thatte_harmoni}, and METIS \citep{2021Msngr_brandl_metis} will provide moderate or high spectral and high spatial resolution IFSs in the near- and mid-infrared, which will be well suited for the detection and characterisation of exoplanets \citep{2021A&A_houlle_harmoni, 2020JATIS_carlomagno_metis_hci}. From space, moderate-resolution IFSs are provided by NIRSPEC-IFU and MIRI-MRS on board of the James Webb Space Telescope \citep{patapis_MM_MIRI}.

These IFSs are inherently limited by their number of available detector pixels. For example, increasing the spectral resolution, spectral bandwidth, FOV, and spatial sampling would lead to an increase in the number of pixels that are required. It is not trivial to choose the optimum values for each of these parameters. Furthermore, in planning observations, it is not obvious if it is better to use the higher spectral resolution mode of an instrument or to have broader wavelength coverage. In this work, we study these trade-offs and the impact of the different parameters on the detection and characterisation of exoplanets. Section \ref{sec:simulations} provides details on the end-to-end simulations and spectral models, followed by the results in Section \ref{sec:sim_results}. We validate some of our simulation results with real observations using VLT/SINFONI in Section \ref{sec:sinfoni}. Finally, in Section \ref{sec:characterisation}  we look at the characterisation potential of such an instrument. Future work will study how these trade-offs change if the main goal of the instrument is to detect exoplanets in reflected light, instead of emission, such as for the future Planetary Camera and Spectrograph (PCS) instrument at the ELT \citep{2021Msngr_Kasper_pcs}.

\section{Simulations}\label{sec:simulations}
    
    \subsection{Instrument model}\label{sec:instrument_sim}
        To study the various trade-offs, we simulated a typical high-contrast imaging instrument on an 8-metre telescope and used VLT/SPHERE \citep{2019A&A_beuzit_sphere} as an example. While our simulations were done for an 8-metre class telescope, most of the trade-offs are also applicable to the integral field spectrographs in the near-infrared of future ELTs, such as HARMONI \citep{2021Msngr_Thatte_harmoni} and PCS \citep{2021Msngr_Kasper_pcs}. Our simulations were done using the \texttt{obsim}\footnote{https://github.com/fjfaggingerauer/obsim} python package for end-to-end simulations (Fagginger-Auer, in prep). This framework allowed for modular Fourier-based simulations of instrument concepts and observations and is build on top of \texttt{HCIPy} \citep{2018SPIE10703E_por_hcipy}.
        
        Our mock instrument consists of the VLT aperture followed by an implementation of the Apodized Pupil Lyot Coronagraph (APLC; \citet{2005ApJ_Soummer_APLC, 2009A&A_Martinez_APLC, 2016ApJ_ndiaye_aplc}), as currently present in VLT/SPHERE. The simulations of the APLC were validated against the \texttt{Coronagraphs} package for python (N'Diaye, private comm.). Instead of doing the full XAO simulations, as is done in \citet{2021A&A_houlle_harmoni}, we used the reconstructed XAO wavefront errors, non-common path aberrations (NCPA), and amplitude errors obtained for SPHERE, as described in Appendix B in \citet{2019A&A_vigan_zelda}. The simulations were run wavelength by wavelength, resulting in a datacube with an image at each wavelength. We ran the simulations for 5000 frames of the reconstructed XAO residuals with a FOV (FOV) of 2'' in diameter and a spectral resolution of 30,000, since it can later be resampled to lower spectral resolutions. The simulations were done in the J, H, and K bands, and the size of the focal plane mask of the APLC was for each band adopted to the values specified in \citet{2019A&A_beuzit_sphere}. 
        
\begin{figure*}[htbp]
            \centering
            \includegraphics[width=\linewidth]{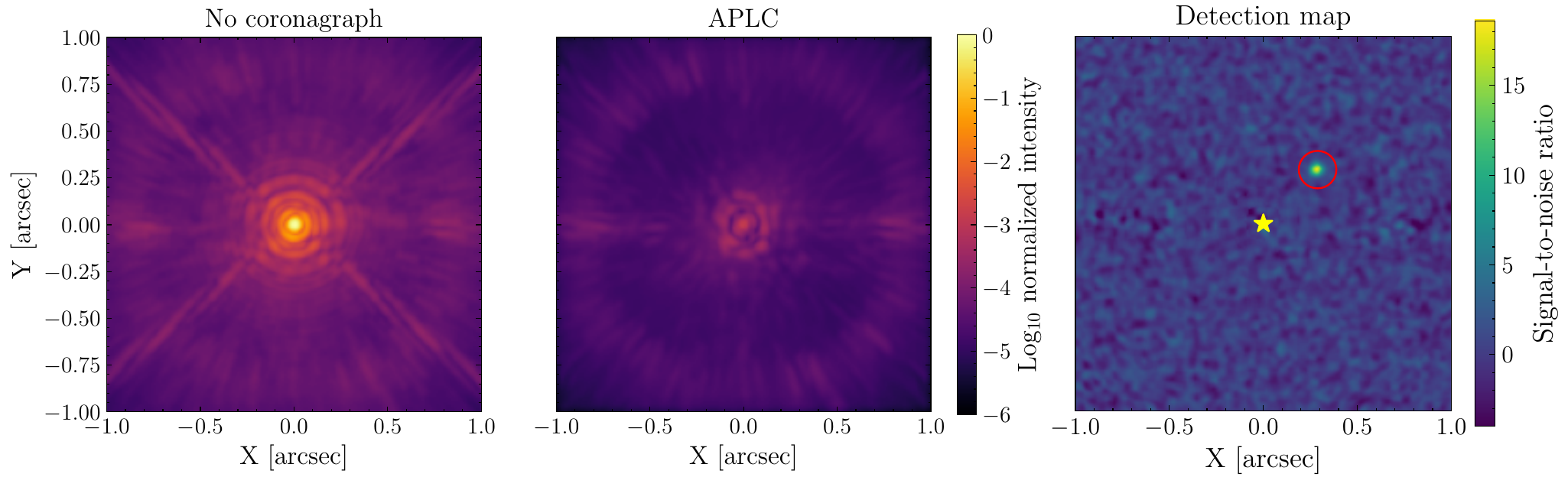}
            \caption{Visualisation of the simulated datacubes. \textbf{Left:} Resulting wavelength summed H-band images from the simulations described in Section \ref{sec:instrument_sim}. \textbf{Middle:} Same but including an implementation of the APLC. \textbf{Right:} Example result of the data analysis. The location of the planet is indicated with a red circle. }
            \label{fig:simulations}
        \end{figure*}

        Since the simulation of the datacubes at this spectral resolution is very time consuming, we opted not to simulate the full cubes for the off-axis planets. Instead, we estimated the average throughput of the planet signal through the APLC as a function of separation in each band. The companion signal was then modelled as the non-coronagraphic datacube shifted to the location of the companion and scaled by the estimated throughput of the coronagraph. Additionally, we assumed a general instrument throughput of 5\%, which is slightly below the estimated optimal throughput for a fiber-based IFS \citep{2020JATIS_haffert_mcifu}. An example of the resulting collapsed noiseless cube with and without the coronagraph in the H-band is shown in Fig. \ref{fig:simulations}.
        
        \subsection{Astrophysical models}
        In this paper, we studied three planets from archetypical, directly imaged planetary systems: $\beta$ Pictoris b \citep{2010Sci_lagrange_beta_pic}, 51 Eridani b \citep{2015Sci_macintosh_51eri}, and HR8799 e \citep{2010Natur_marois_hr8799e}. The assumed properties of these planets as used in the simulations are specified in Table \ref{tab:planets}. Furthermore, we assumed solar metallicity and elemental abundances for all of them and a radial velocity of 20 km/s between the target and observer. The input emission spectra for these planets were obtained from the BT-SETTL grid of atmosphere models \citep{2014IAUS_Allard_btsettl}, while the PHOENIX stellar models \citep{2013A&A_husser_phoenix} were used for the stellar spectra.
        
        \subsection{Mock observations}\label{sec:mock_obs}
        We used the atmospheric transmission and sky emission from ESO's SkyCalc tool \citep{2012A&A_Noll_skycalc} for an airmass of 1.1 and precipitable water vapour of 2.5 mm. We did not consider thermal emission from the instrument itself, which may start to contribute in the K-band. The baseline simulations consist of 60 times 60 second exposures, and the flux in each wavelength bin in the cubes was scaled according to the expected number of received photons from the spectra and instrument properties.
    
        \begin{table*}[]
            \centering
            \caption{Assumed parameters of the planetary systems used for the simulations throughout this paper. References: $\beta$ Pic b: \citet{2013A&A_bonnefoy_beta_pic_b}, \citet{2014Natur_snellen_beta_pic_crires}; HR8799e: \citet{2019A&A_gravity_hr8799e, 2020A&A_molliere_hr8799, Wang_2021AJ_hr8799_kpic}; and 51 Eri b: \citet{2017A&A_samland_51eri, 2017AJ_rajan_51eri}}
            \label{tab:planets}
            \def\arraystretch{1.3}
            \begin{tabular}{|l|c|c|c|c|c|c|c|c|}
            \hline
            Planet                           & $R_*$                             & $T_*$   & Distance & $R_p$            & $T_p$   & log(g) & Separation & Planet vsin(i) \\ \hline
            \multicolumn{1}{|l|}{$\beta$ Pic b} & 1.80 $R_{\odot}$  & 8100 K & 19.4 pc & 1.36 $R_{\textrm{jup}}$ & 1700 K & 4.0  & 0.30'' & 25 km/s \\
            \multicolumn{1}{|l|}{HR8799 e}       & 1.34 $R_{\odot}$ & 7400 K & 40.88 pc & 1.12 $R_{\textrm{jup}}$ & 1200 K & 4.0  & 0.40''  & 15 km/s      \\
            \multicolumn{1}{|l|}{51 Eri b}      & 1.45 $R_{\odot}$ & 7200 K & 29.4 pc  & 1.10 $R_{\textrm{jup}}$  & 700 K  & 4.0  & 0.45''  & 10 km/s      
            \\\hline
            \end{tabular}
        \end{table*}

    \subsection{Molecule mapping}
    Detecting a planet in the resulting datacubes can be done using the molecule mapping technique developed by \citet{2018A&A_hoeijmakers_mm}. In this technique, the stellar and telluric contamination is removed from each spaxel, leaving the planet signal with noise. Subsequently, a matched filter or cross-correlation is applied to the residuals. If there is uncorrelated Gaussian noise, the S/N ratio of a matched filter is given by \citep{2017ApJ_ruffio_fmmf, 2021A&A_houlle_harmoni}:
       \begin{equation}
        \label{eq:mf_snr}
        S/N = \frac{\sum_i^N s_i t_i/\sigma_i^2}{\sqrt{\sum_i t_i^2/\sigma_i^2}},
    \end{equation}
    where $s_i$ is the observed data, $t_i$ the template, $\sigma_i$ the uncertainties, and $N$ the total number of data points. These data points, indexed by i, can be pixels spanning both the spectral domain (the planetary spectrum) and spatial domain (the shape of the Point Spread Function, or PSF).
    For the simulations, we knew the noise of each data point exactly, allowing us to directly apply Eq. \ref{eq:mf_snr}. The noise of a data point $\sigma_i$ is given by:

        \begin{equation}
       \sigma_i = \sqrt{F^*_i + F^p_i + F^{bg}_i + \textrm{RON}^2}.
    \end{equation}
    Here, $F^*_i$, $F^p_i$, and $F^{bg}$ are respectively the stellar, planet, and background contribution to the noise at the considered spaxel and spectral bin. These contributions to the noise were obtained from the simulated datacubes described in Section \ref{sec:instrument_sim}. \textrm{The value of RON} is the total readout noise of each spectral bin. For the baseline simulations, we assumed each spectral bin effectively covers two physical pixels on the detector and has a readout noise of 1 e$^-$ rms/pixel/read.
    
    We experimented with adding noise to the datacubes and then doing the full data analysis following \citet{2018A&A_hoeijmakers_mm} to estimate the S/N, and we found results equivalent  to those of Eq. \ref{eq:mf_snr}. To have consistency across the different instrument configurations and avoid dependency on the choice of the values of parameters of the algorithm, we have opted to only present the analytical results from Eq. \ref{eq:mf_snr}, just as was done in \citet{2021A&A_otten_hirise}. 
    
    The matched filter templates $t_i$ used throughout this paper consist of the BT-SETTL spectra and molecular templates generated using \texttt{petitRADTRANS} \citep{2019A&A_molliere_prt}, convolved to the appropriate spectral resolution. Molecule mapping generally results in the loss of the continuum, and we therefore high-pass filtered these templates. This was done by convolving the original templates with a Gaussian kernel with a standard deviation of 0.01 $\mu$m and subtracting this from the original spectrum. Examples of high-pass filtered templates are shown in Appendix \ref{app:models}. For the spatial part of the matched filter, we used a Gaussian function with an appropriate full-width at half maximum (FWHM) given the wavelength.

\section{Trade-off results}\label{sec:sim_results}
    In this section, we study how different properties of the instrument impact its ability to detect new planets. This is be done based on analytical considerations and the simulations described in Section \ref{sec:simulations}. The studied properties are the spectral resolution, spectral bandwidth, observing band, and FOV.

    \subsection{Spectral resolution}\label{sec:sim_R}
     The impact of the spectral resolution on the S/N that can be achieved is not trivial. Increasing the spectral resolution leads to more data points for the matched filter ($N \propto R$) but decreases the S/N per spectral bin. 
     Furthermore, the observed spectrum also changes as a function of spectral resolution, as individual lines may start to be resolved and become deeper.
        
        \subsubsection{Power spectral density}\label{sec:psd}
\begin{figure}[htbp]
            \centering
            \includegraphics[width=\linewidth]{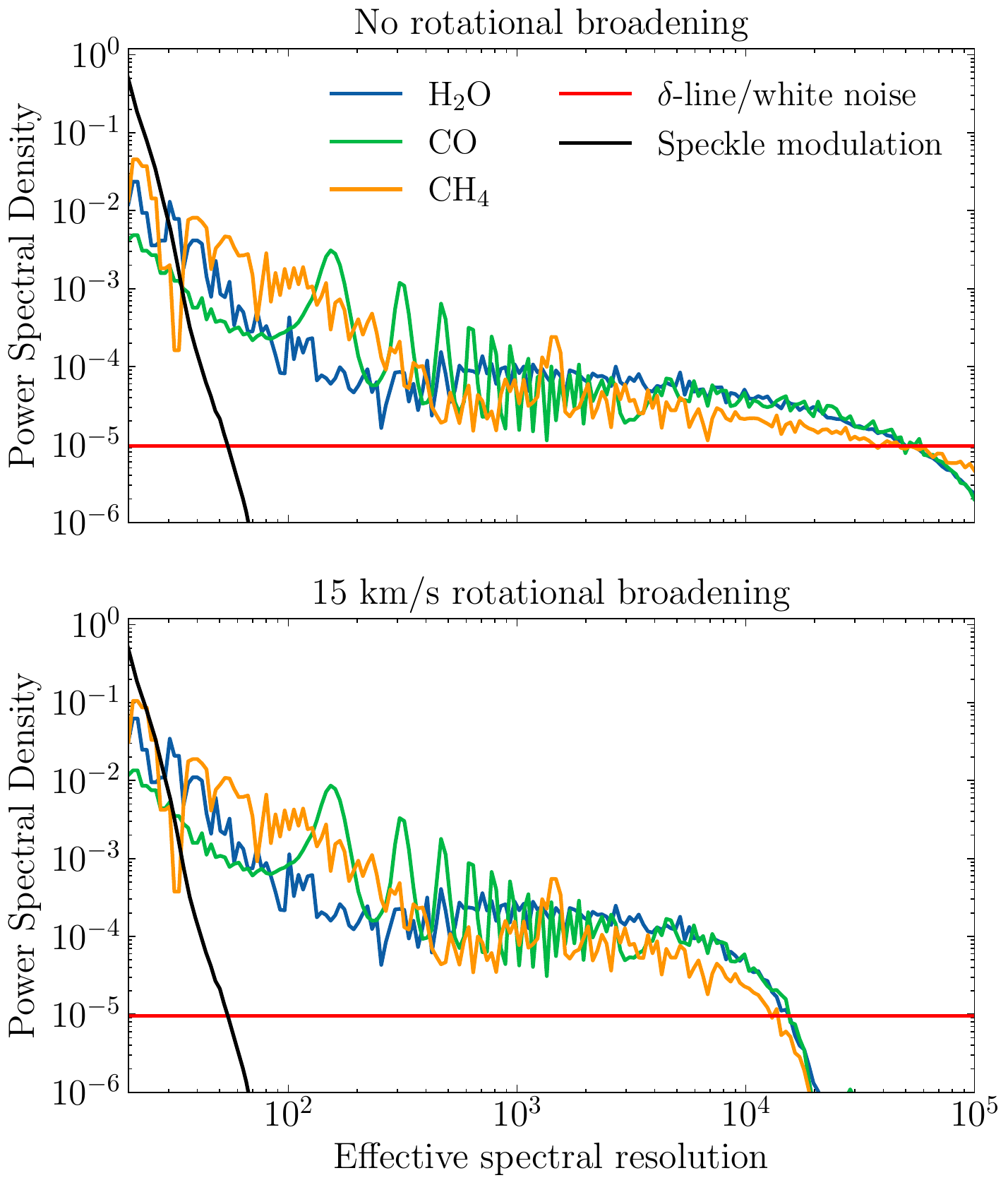}
           \caption{Normalised PSD of different contributions to the observed spectrum between 1 and 2.5 $\mu$m. Here, we assumed a planet with an effective temperature of 1200 K and log(g) of 4.0. The panels show the results without rotational broadening (top) and with rotational broadening of $v_{\textrm{rot}}\sin i = 15$ km/s (bottom).}
            \label{fig:psd}
        \end{figure}

        We analysed the strength of the features that are resolved at a specific spectral resolution by looking at the power spectral density (PSD) of the planet spectrum $t_p$. The PSD of a sequence $t_p(x)$ defined over a range $L = x_{max}- x_{min}$ can be approximated by:
        \begin{equation}
            \begin{split}
            \textrm{PSD}[t_{p}(x)](f) &= \frac{1}{L}|\mathcal{F}[t_p (x)]|^2 \\
            &\approx \frac{1}{L}\left|\int_{x_{min}}^{x_{max}} t_p(x) e^{-i2\pi fx} dx \right|^2,
            \end{split}
        \end{equation}
        where $\mathcal{F}$ is the Fourier transform operator, $t_p$ the considered planetary spectrum, and $f$ the frequency of the corresponding fluctuation. This PSD can also be expressed in terms of the effective spectral resolution $R'$ using $R' = \lambda/(2\Delta \lambda')$. We did this by sampling the planet spectrum logarithmically in wavelength (i.e. $x=\log \lambda$), keeping the spectral resolution constant across the wavelength range. In this case, the conjugate variable in the Fourier domain is $f\approx\lambda/\Delta \lambda'= 2R'$ (see Appendix \ref{app:derivations}). The factor two was added to account for the fact that we needed to Nyquist sample the spectrum to see fluctuations of a specific frequency. It is important to note that $\Delta \lambda'$ and $R'$ refer to the period and frequency of a fluctuation respectively and not the actual resolution element of the instrument. Using the definition of the PSD from above, we obtained:
        \begin{equation}\label{eq:planet_psd_def}
            \begin{split}
            P(R') &\equiv \textrm{PSD}[t_p(\log\lambda)](f/2)\\
            &  \approx \frac{1}{L}\left|\int_{\log \lambda_{min}}^{\log \lambda_{max}} t_p(\log \lambda) e^{-i2\pi R'\log \lambda} d\log \lambda \right|^2,
            \end{split}
        \end{equation}
        with $L=\log \lambda_{max} - \log \lambda_{min}$. The normalised PSD of different contributions to the observed spectrum between 1 and 2.5 $\mu$m is shown in Fig. \ref{fig:psd} as a function of effective spectral resolution. The PSDs were estimated using Welch's method and smoothed for visibility purposes. The planet contributions presented here are for a HR8799e-like planet and are shown both with and without rotational broadening. The PSD's of the other reference targets are shown in Appendix \ref{app:psd}. We also show the PSD for a single infinitely thin spectral line ($\delta$-line) and white noise. Finally, because of the wavelength dependence of the PSF, stellar speckles shift outwards for increasing wavelengths. This imposes a low-order modulation to the stellar contribution to each spaxel and depends on the separation of the source and the present wavefront aberrations. Figure \ref{fig:psd} also shows the average PSD of this speckle modulation, which we calculated in a square area between 2 and 7 $\lambda/D$ from the simulated coronagraphic datacubes from Section \ref{sec:simulations}.
        
         Figure \ref{fig:psd} shows that the modulation of stellar speckles is a low-order effect and has the most power at low spectral resolutions. The power of these fluctuations very quickly decreases at higher spectral resolution. This illustrates that applying a spectral high-pass filter is indeed very effective at removing stellar contamination, as it leaves only narrow spectral features such as stellar and telluric lines. Since these are shared between all of the spaxels, they can often be removed using, for example, a principal component analysis \citep[PCA; e.g. ][]{2018A&A_hoeijmakers_mm, Ruffio_2021AJ_HR8799_deep}.
         
        The PSDs of the molecular templates are more complex. Figure \ref{fig:psd} shows that the power generally decreases for higher spectral resolutions. This is the result of the band structure of the absorption spectra. Furthermore, at very high resolutions ($R>50,000$), we also observed decreasing power due to the intrinsic broadening of the individual ro-vibrational lines. We also found that, mostly for CO and CH$_4$, there are clear peaks in the power spectrum. This is because of the ro-vibrational structure of the absorption features, which can have a specific periodicity and are therefore resolved at a specific spectral resolution. This is especially prominent for CO, which has a sawtooth structure in its bandheads. Finally, both a $\delta$-line and white noise have a flat PSD and are the dominating contributors at the highest spectral resolutions. While the PSD of the molecular features depends slightly on the parameters of the planet, such as the temperature structure, the general trend is the same, as can be seen in Appendix \ref{app:psd}.

        \subsubsection{Signal-to-noise ratio approximation}
        Next, we discuss the relation between the S/N and spectral resolution of the instrument. If we assume that we have a perfect match between the template $t$ and the observed data $s$ with some uncorrelated Gaussian noise $n_i = \mathcal{N}(0, \sigma_i)$ we have
        \begin{equation}
            s_i = t_i + n_i,
        \end{equation}
        simplifying Eq. \ref{eq:mf_snr} to:
        \begin{equation}
            S/N = \frac{\sum_i^N t_i^2/\sigma_i^2}{\sqrt{\sum_i^N t_i^2/\sigma_i^2}} = \sqrt{\sum_i^N \frac{t_i^2}{\sigma_i^2}}.
        \end{equation}
        
        For simplicity, we assumed that the planet flux is concentrated in a single spaxel with $M$ spectral bins. Further assuming that the noise in each spectral bin is roughly the same ($\sigma_i \approx \sigma$), we could approximate this with:
        
        \begin{equation}\label{eq:snr_var}
            S/N \approx \frac{1}{\sigma}\sqrt{\sum_i^M t_i^2} = \frac{1}{\sigma}\sqrt{M <t^2>},
        \end{equation}

        where $<t^2>$ denotes the average of $t^2$. From this equation, we could show the following relation between the matched filter S/N and the PSD (see Appendix \ref{app:derivations} for the derivation):
        \begin{equation}\label{eq:snr_psd}
            S/N \approx \frac{2\sqrt{M}}{\sigma} \sqrt{\int_0^\infty |H(R')|^2P(R') dR'}.
        \end{equation}  
        Here, $P(R')$ is the PSD of the planet spectrum as discussed in section \ref{sec:psd}, and $H(R')$ is the transfer function describing the impact of the spectrograph, detector, and data reduction on the observed spectrum. Assuming an ideal Nyquist-sampled spectrograph that acts as an ideal low-pass filter up to the Nyquist frequency set by the spectral resolution $R$ and that the data reduction functions as an ideal high-pass filter, removing all features below $R'=R_{\textrm{min}}$, we obtained the following equation (see Appendix \ref{app:derivations} for details):
        \begin{equation}\label{eq:snr_psd_ideal_lsf}
            S/N  \approx \frac{2\sqrt{2B}}{\sigma\sqrt{R}} \sqrt{\int_{R_{\textrm{min}}}^{R} P(R') dR'},
        \end{equation}

       where we have used $M=2BR$ for Nyquist-sampled spectra and with $B= (\lambda_{\textrm{max}} - \lambda_{\textrm{min}})/\lambda_{\textrm{central}}$ as the relative spectral bandwidth. While the true line spread function and detector sampling of the instrument may differ and are strongly dependent on the spectrograph design, this is beyond the scope of this work. A comparison of Equations \ref{eq:snr_psd} and \ref{eq:snr_psd_ideal_lsf} to full numerical simulations for a specific instrument configuration is shown in Appendix \ref{app:equation_validation}.
        
        \subsubsection{Noise regimes}
        The relation between the noise per bin $\sigma$ and the spectral resolution depends on the noise regime we are in. For photon noise limited observations, we have $\sigma(R) \propto 1/\sqrt{R}$. This is because the stellar flux is distributed over $R$ times more pixels, and we thus have $\sqrt{R}$ less shot noise per pixel. In the detector noise limited case, we have $\sigma(R)=\textrm{constant}$, as the noise per pixel stays the same. 
        
        The situation in the speckle-limited regime is a little more complex. Here we defined the speckle noise as all spectral fluctuations as a result of phase aberrations in the pupil plane. While in reality, these speckles can also contain high-frequency fluctuations due to imperfect telluric correction and stellar lines, this is highly dependent on the data analysis and planetary spectrum, among other factors. The amplitude of the speckle modulation decreases by a factor of $R$\, in this case as well because the stellar flux is distributed over $R$ times more pixels. However, we pick up more power from the speckle fluctuations, as seen from the PSD in Fig. \ref{fig:psd}. The noise due to speckles can be approximated by:
        \begin{equation}
        \begin{split}
            \sigma_{\textrm{speckle}}(R) \propto \frac{1}{R}\sqrt{\int_{R_{\textrm{min}}}^{R}P_{\textrm{speckle}}(R') dR'}.
            \end{split}
        \end{equation}
        Since most of the power is at low effective spectral resolutions, the integral is constant after $R\gtrsim100$, and we could thus approximate it with $\sigma(R)\propto 1/R$ for higher spectral resolutions. The data reduction filter can, in principle, be designed in such a way as to minimze $\sigma_{\textrm{speckle}}$ while minimally impacting the planet signal. Even though speckle noise is by definition correlated and thus invalidates Eq. \ref{eq:mf_snr}, we have numerically validated that the equation is still a good approximation of the classical cross-correlation S/N \citep[e.g.][see Appendix C)]{2013MNRAS_birkby_water_hd189, 2018A&A_hoeijmakers_mm}.
        
        \begin{table}[]
            \centering
            \def\arraystretch{1.8}
            \begin{tabular}{l|c|c|c}
                 Noise regime & $\sigma(R) \propto$ &  S/N $\propto$ & $\delta$-line S/N$\propto$ \\\hline \hline
                 Speckle noise & $1/R$ & $\mathcal{T}(R)\sqrt{R}$ & $R$\\\hline
                 Photon noise & $1/\sqrt{R}$ & $\mathcal{T}(R)$ & $\sqrt{R}$ \\\hline
                 Detector noise & constant & $\mathcal{T}(R)/\sqrt{R}$ & constant\\
            \end{tabular}
            \caption{Summary of the relation between the S/N of a matched filter and the spectral resolution in the three different noise regimes with $\mathcal{T}(R)= \sqrt{\int_{R_{\textrm{min}}}^{R} P(R') dR'}$.}
            \label{tab:snr_noise_regimes}
        \end{table}
        
         \begin{figure*}[htbp]
            \centering
            \includegraphics[width=\linewidth]{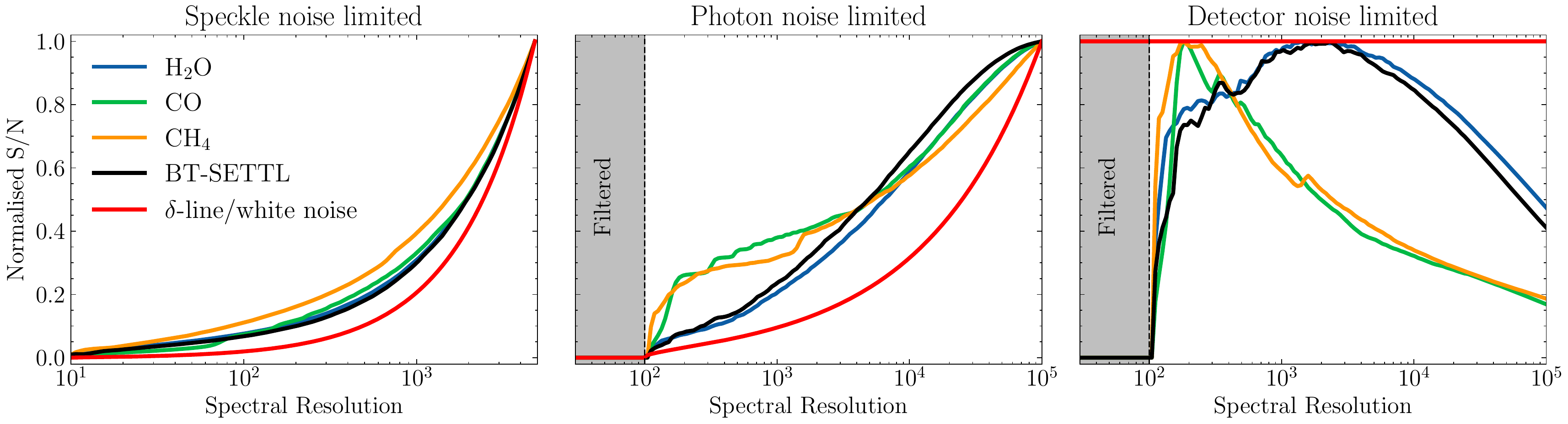}
           \caption{Normalised S/N ratio as a function of spectral resolution for different contributions as calculated with Eq. \ref{eq:snr_psd}. From left to right are the speckle noise, photon noise, and detector noise limited regimes. A high-pass filter was applied to move from the speckle-limited regime to the photon and detector-limited regime. A planet with an effective temperature of 1200K and no rotational broadening is shown here.}
            \label{fig:snr_vs_R}
        \end{figure*}

        \subsubsection{Signal-to-noise curves}\label{sec:sim_snr_vs_R}
        The relation between the S/N and the spectral resolution for different noise regimes, as described in the previous section, is summarised in Table \ref{tab:snr_noise_regimes} and plotted in Fig. \ref{fig:snr_vs_R}. There is a factor $1/\sqrt{R}$ difference between the speckle-limited regime and the photon noise limited regime and another $1/\sqrt{R}$ factor to the detector noise limited regime. This shows that increasing the spectral resolution is the most significant in the speckle-limited regime, as it allows us to distinguish between speckles and planetary features. To move out of the speckle-limited regime, one can apply a spectral high-pass filter, which is effectively also done in molecule mapping \citep{2018A&A_hoeijmakers_mm}. In this case, we set the filter cutoff at $R'=100$. In the photon-limited regime and a spectrum consisting of white noise or a $\delta$-line with a flat PSD, we have $S/N\propto \sqrt{R}$, which is the empirical relation assumed by \citet{2018A&A_hoeijmakers_mm}. Molecular absorption spectra have larger gains in S/N at lower spectral resolutions and less gain at higher spectral resolutions, as compared to a series of $\delta$-lines, which is expected from the PSD. This effect may further be strengthened by rotational broadening of the signal, which removes power at high spectral resolution. This means that moderate-resolution spectroscopy is already powerful for targeting molecular absorption features, while higher spectral resolution is preferable for studying individual or narrow emission and absorption lines. Furthermore, CO and CH$_4$ have relatively more features at spectral resolutions less than 1000 due to their prominent band structure, while the S/N for water increases at a somewhat higher spectral resolution. An important caveat is that we have assumed perfect removal of the stellar spectrum and telluric contamination, which may be harder at a lower spectral resolution.
         
         Finally, in the detector-limited regime, we observed that there is an optimal spectral resolution for the molecular templates, after which the S/N decreases again. This optimum is reached at $R\sim 2,000$ for water and the BT-SETTL spectrum, which is dominated by water absorption at this temperature. For CO and methane, this optimum is already reached at $R\sim 100$, which is again the result of the strong band structure of absorption features of these molecules, as opposed to the more distributed water lines.

    \subsection{Spectral bandwidth}\label{sec:sim_R_vs_B}
\begin{figure}[htbp]
            \centering
            \includegraphics[width=\linewidth]{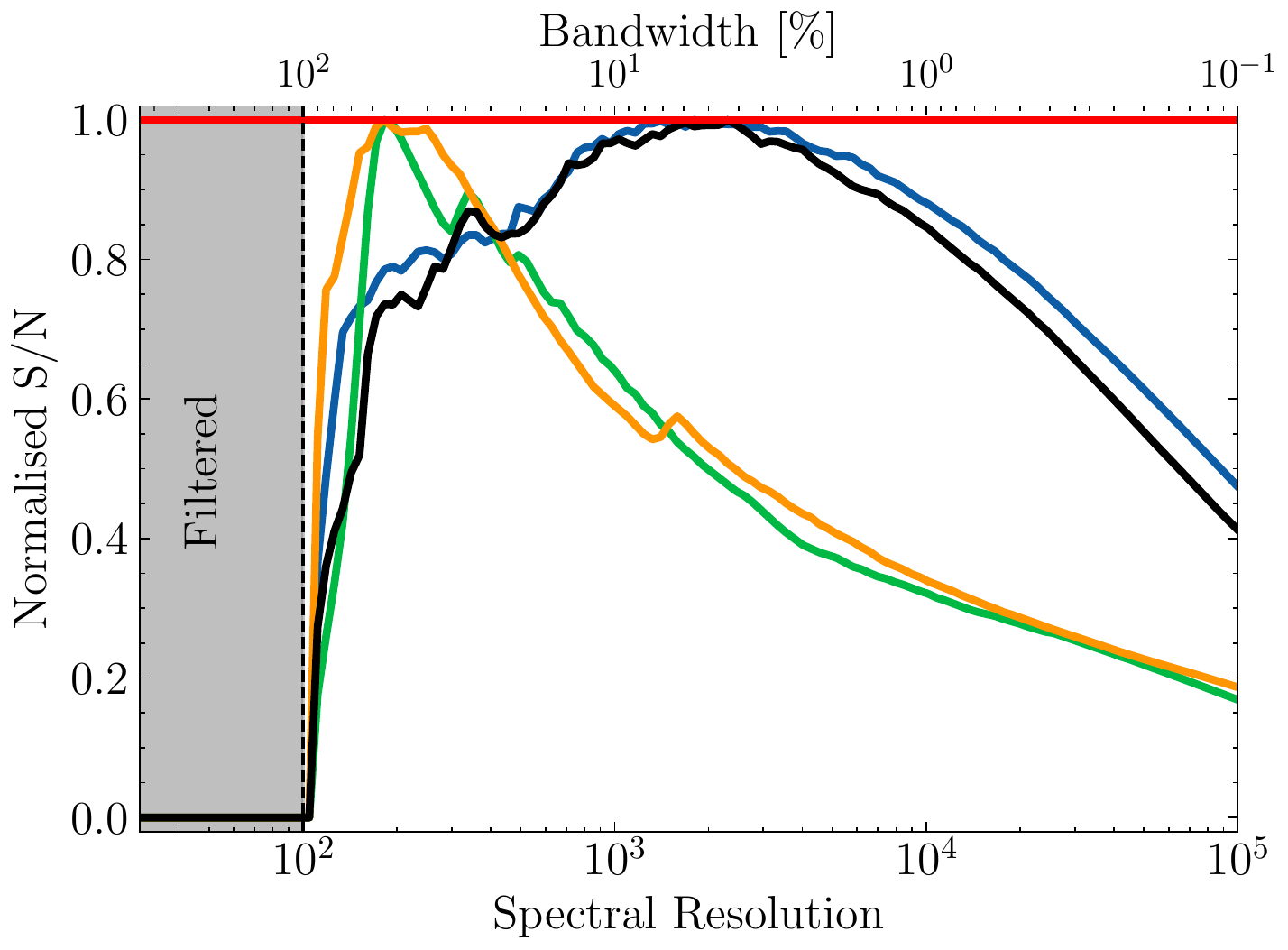}
           \caption{Trade-off between spectral resolution and bandwidth showing the normalised S/N in the case where the total number of spectral bins is constant. Planet models with an effective temperature of 1200 K were used.}
            \label{fig:R_vs_B}
        \end{figure}

        For a fixed detector size, plate scale, and FOV, there are a fixed number of available spectral bins per spaxel. Thus, this raises the question as to whether it is better to increase the spectral resolution or the spectral bandwidth. Here, we assumed that we are increasing the bandwidth within the absorption features of the targeted species and thus that we capture more lines. While this assumption is only valid for CO over a small wavelength range, it is more reasonable for water, which is the dominant contributor to the spectra of the planets specified in Table \ref{tab:planets}. The assumption and the wavelength range in which different molecules have absorption features is discussed in Section \ref{sec:snr_vs_wavelength}. In this simplified case, it can be seen from Eq. \ref{eq:snr_psd} that:
        \begin{equation}
            S/N(B) \propto \sqrt{B}.
        \end{equation}
        Since the product of $B \times R$ needs to be constant to not increase the number of required pixels, it is thus preferable to increase the spectral resolution if the increase in the S/N is steeper than $\sqrt{R}$. Following our calculations from Section \ref{sec:sim_R} and assuming that the globally calculated power spectrum is applicable locally, we were able to calculate this trade-off. The result for a planet with an effective temperature of $\sim$ 1200K in the photon-limited regime is shown in Fig. \ref{fig:R_vs_B}, assuming there is a total of 200 spectral bins per spaxel. We ended up with the same S/N curve as in the detector-limited regime from Fig. \ref{fig:snr_vs_R}. If the spectrum consists of $\delta$-lines, it does not matter whether we increase the spectral resolution or the bandwidth, as the S/N is proportional to $\sqrt{B\times R}$. However, for water and a BT-SETTL spectrum, we found that it is best to increase the spectral resolution until $\sim$2,000 and after that to first maximise the bandwidth. If the goal is to detect the presence of methane or carbon monoxide, the optimum spectral resolution is even lower. However, since these molecules have very specific wavelength ranges in which they have absorption features, as is shown in Section \ref{sec:snr_vs_wavelength}, it is usually not beneficial to increase the bandwidth beyond these absorption bands.
        
    \begin{figure*}[htbp]
        \centering
        \includegraphics[width=\linewidth]{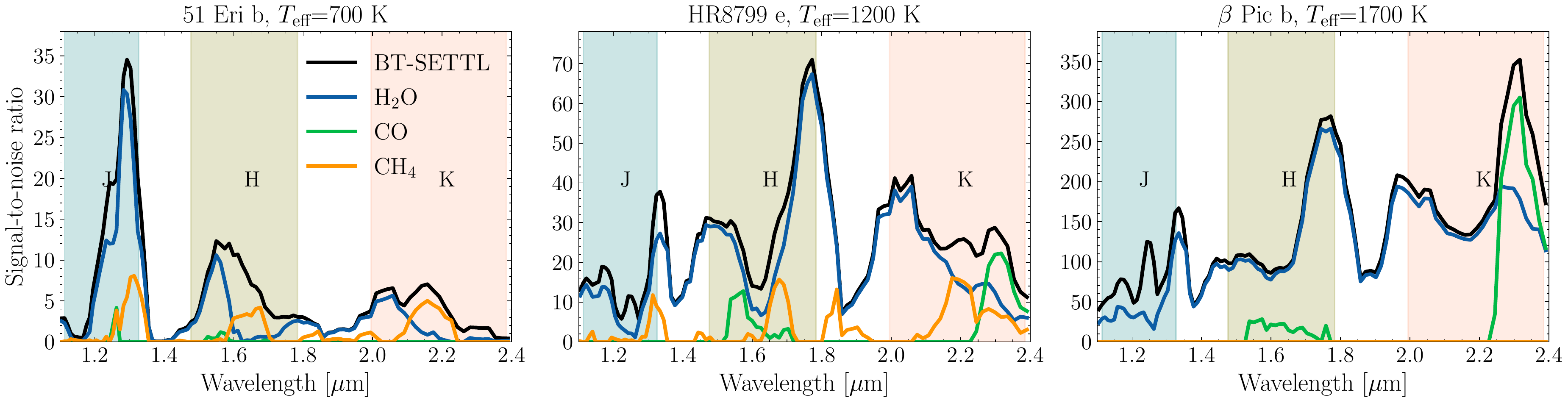}
        \caption{Matched filter S/N between a BT-SETTL spectral model and molecular templates as a function of wavelength, assuming a 3\% spectral bandwidth, a spectral resolution of 5000, and photon-limited observations. This is shown for three typical systems (from left to right): 51 Eri b, HR8799 e, and $\beta$ Pic b.}
        \label{fig:snr_vs_wl}
    \end{figure*}
        
        \subsection{Central wavelength}\label{sec:snr_vs_wavelength}
        In the previous section we assumed that the molecular absorption features are evenly distributed across the wavelength range and the noise is constant. However, this is not realistic, and we have to choose the optimal wavelength range for the instrument. We considered ground-based observations and focussed on the J, H, and K bands in the near-infrared. In these bands, the spectra of warm gas giants exhibit significant molecular absorption, while the sky background remains manageable. These are also the bands that are predominantly used in current high-contrast imaging instruments. There are multiple effects that impact the S/N that can be achieved as a function of wavelength. Firstly, the emitted flux of the planet is wavelength dependent, and molecular absorption bands are present at specific wavelength ranges. Secondly, a higher Strehl ratio, and more starlight suppression, can be achieved at longer wavelengths. On the other hand, the separation between the star and planet is less $\lambda$/D at these longer wavelengths, which could mean worse performance of the coronagraph. Finally, the sky background starts to increase in the K-band. 
        
        Our framework presented in Section \ref{sec:simulations} considers all these effects. We used the simulated datacubes for the mock observation settings described in Section \ref{sec:mock_obs} and the three systems from Table \ref{tab:planets}. We calculated the matched filter S/N for a 3\% bandwidth around each wavelength bin using Eq. \ref{eq:mf_snr}, assuming a spectral resolution of 5,000. This was done for four different templates, a BT-SETTL model of the appropriate effective temperature and the three most prominent molecular absorbers in this wavelength range, namely, H$_2$O, CO, and CH$_4$. The results are shown in Fig. \ref{fig:snr_vs_wl}. We observed that T-type companions, such as 51 Eri b, are best detectable in the J-band and are dominated by water and methane absorption. On the other hand, L-type companions, such as $\beta$ Pic b, are best observed in the K-band, with features of both CO and water. The H-band is well suited to detect both types of planets and has mainly water features and some methane for cooler planets. Planets with effective temperatures between these two have dominant spectral features in all three bands. We note that the highest S/N can often be achieved at the edges of the bands, especially for water. For example, there is a sharp feature at the end of the J-band, which is due to the start of the 1.4 $\mu$m water absorption band. Going to higher spectral resolutions may allow us to observe more efficiently at the edges of the bands, especially if there is a significant Doppler shift between the target and the Earth. When looking back at the assumption in Sec. \ref{sec:sim_R_vs_B}, we observed that features are not evenly distributed across the wavelength range, especially for cooler planets and when targeting CO and CH$_4$. This will shift the optimum of the trade-off to higher spectral resolutions and a smaller bandwidth. The amount by which this is shifted is strongly dependent on the number of available spectral bins and the central wavelength.

    \subsection{Field of view}
    Detecting planets using molecule mapping is most beneficial in the speckle-limited regime. This is because the high-pass filter is very effective at removing stellar speckles while leaving the high-order features in the planet spectrum intact. However, it generally results in the loss of the planet continuum. Farther away from the star, the gain in contrast from the speckle removal might be smaller than the loss of contrast from the removal of the planet continuum. In such cases, classical speckle removal post-processing techniques that keep the continuum (e.g. ADI) are expected to perform better. This limits the required FOV where molecule mapping is useful for increasing the detection capabilities of the instrument. The point at which this occurs depends on many different factors. To illustrate this, we considered an instrument with a 2k$\times$2k detector and considered the mock observations as described in Section \ref{sec:mock_obs}. We assume that our instrument effectively uses 10\% of the detector because of the placement of the spectra on the detector and is Nyquist sampled in the spatial and spectral dimensions. By increasing the FOV, we have fewer available detector pixels per spaxel. For example, for a FOV of 1.0" in diameter in the H-band, there are about 800 pixels/spaxel, while for an FOV of 0.5", there are about 320 pixels/spaxel. This means we can increase the spectral resolution or bandwidth by decreasing the FOV, improving the gain that can be achieved with molecule mapping. We constructed contrast curves by considering ten different position angles for the planet at each separation and calculating the S/N according to Eq. \ref{eq:mf_snr}. If the planet is detected at an S/N higher than five, it is considered detected. If not, the contrast between the planet and the star is decreased, and the process is repeated. The resulting contrast curves for different FOVs and spectral resolutions is shown in Fig. \ref{fig:contrast_curves}. For simplicity, the wavelength coverage is centred at the mean wavelength of each band. We observed that a deeper contrast can be obtained for 51 Eri b than $\beta$ Pic b, which is because cooler planets generally have more spectral features with respect to the continuum. We also saw that a higher spectral resolution increases the achievable contrast, at the cost of a smaller discovery space.
    
    As we wanted to compare the obtained contrast curves to the performance of the ADI,  we divided up the available data of the reconstructed XAO residuals into ten sequential sets and simulated the resulting broadband images. A model of the PSF was then obtained by taking the median of these images. This model was subsequently subtracted from all the individual images, which were summed to obtain the residual image. The residual speckle noise was then estimated by radially computing the standard deviation. We then corrected for the ADI throughput using injection and recovery, assuming a total field rotation of 40 degrees. This was a simplified simulation of the ADI performance, and it is likely overestimated, as we did not consider evolving the NCPA, for example. Still, it demonstrates the point that at a certain separation, the loss of continuum in molecule mapping degrades the achievable contrast, unlike ADI, which retains the continuum. These simulations show that in this specific case, it is not beneficial to go beyond a separation of $\sim$0.5" for the simulated 51 Eri b observations and beyond $\sim$0.3" for the simulated $\beta$ Pic b observations. An important note is that for molecule mapping, assuming negligible detector noise, we have $S/N \propto \sqrt{t_{\textrm{exp}}}$, and we can thus increase the contrast by integrating longer. On the other hand, this is not necessarily the case for ADI, as we may be limited by quasi-static speckles that do not average over time \citep{2022A&AVigan_temporal}. On the other hand, ADI is less affected by phenomenon that dampen spectral features, such as clouds in the exoplanet's atmosphere \citep{2020A&A_molliere_hr8799} or dust extinction \citep{Cugno2021A&A_MM_PDS70}. Finally, we note that both methods are not mutually exclusive, as a spectral matched filter can in principle still be applied after using ADI for speckle removal.
    
         \begin{figure*}[htbp]
            \centering
            \includegraphics[width=0.49\linewidth]{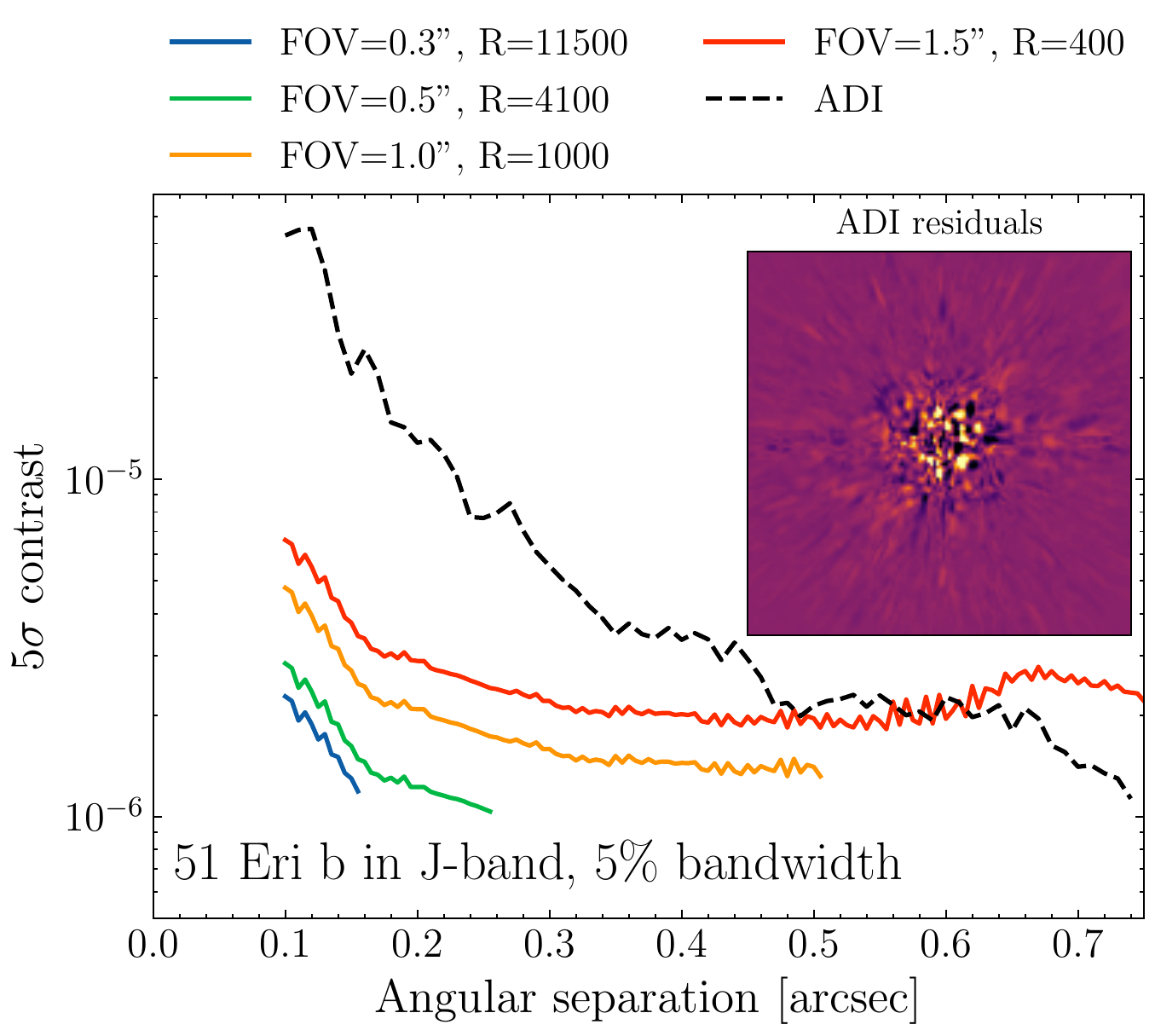}
            \includegraphics[width=0.49\linewidth]{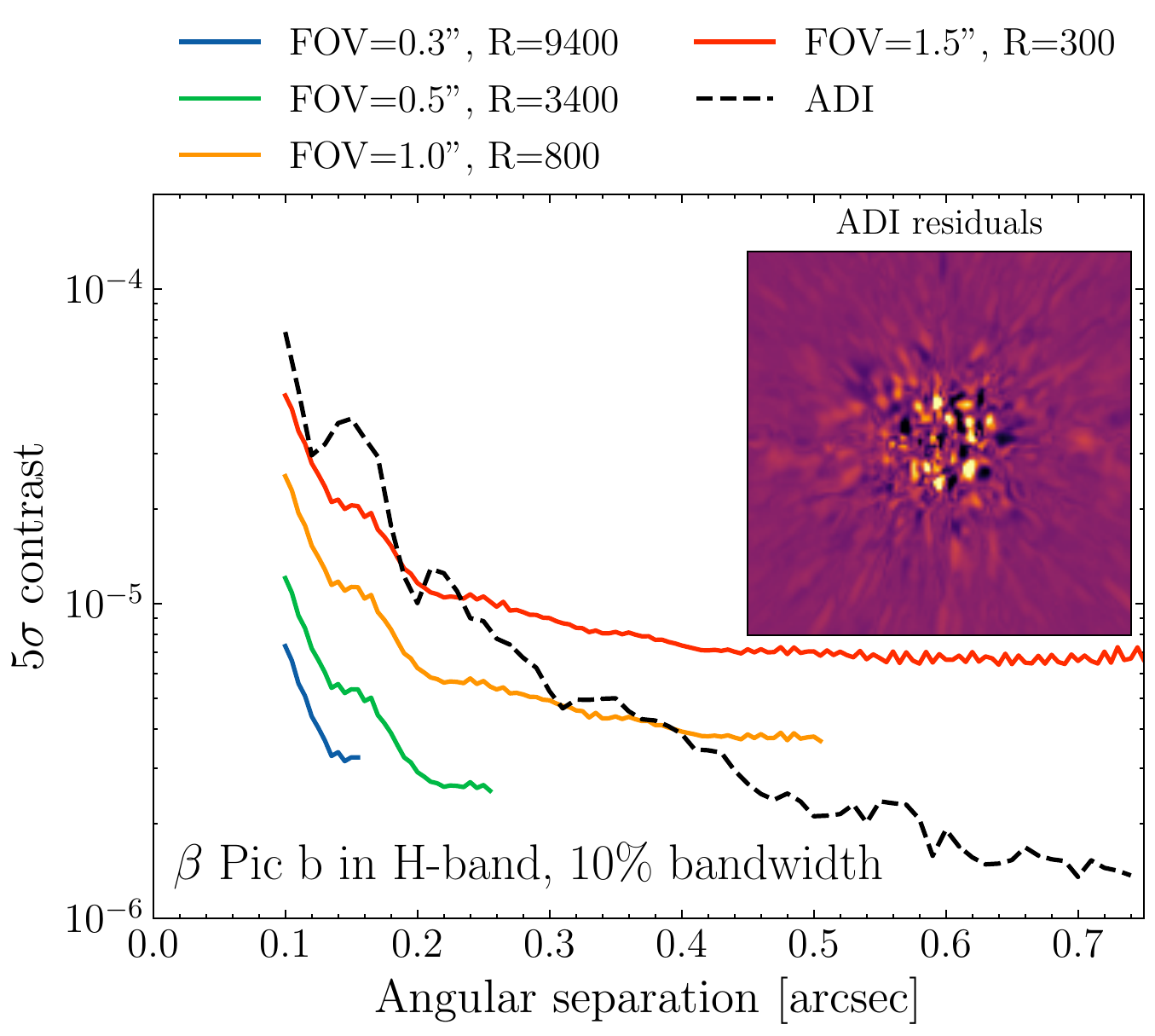}
           
           \caption{Contrast curves with molecule mapping from the simulated observations. Different lines are for different combinations of spectral resolution and FOVs, and all use the same number of detector pixels. The inset shows the residuals after applying ADI on the mock observations. \textbf{Left:} 51 Eri b in the J-band with a 5\% spectral bandwidth. \textbf{Right:} $\beta$ Pic b in the H-band with a 10\% bandwidth. }
            \label{fig:contrast_curves}
        \end{figure*}

        \begin{figure*}
         \includegraphics[width=0.49\linewidth]{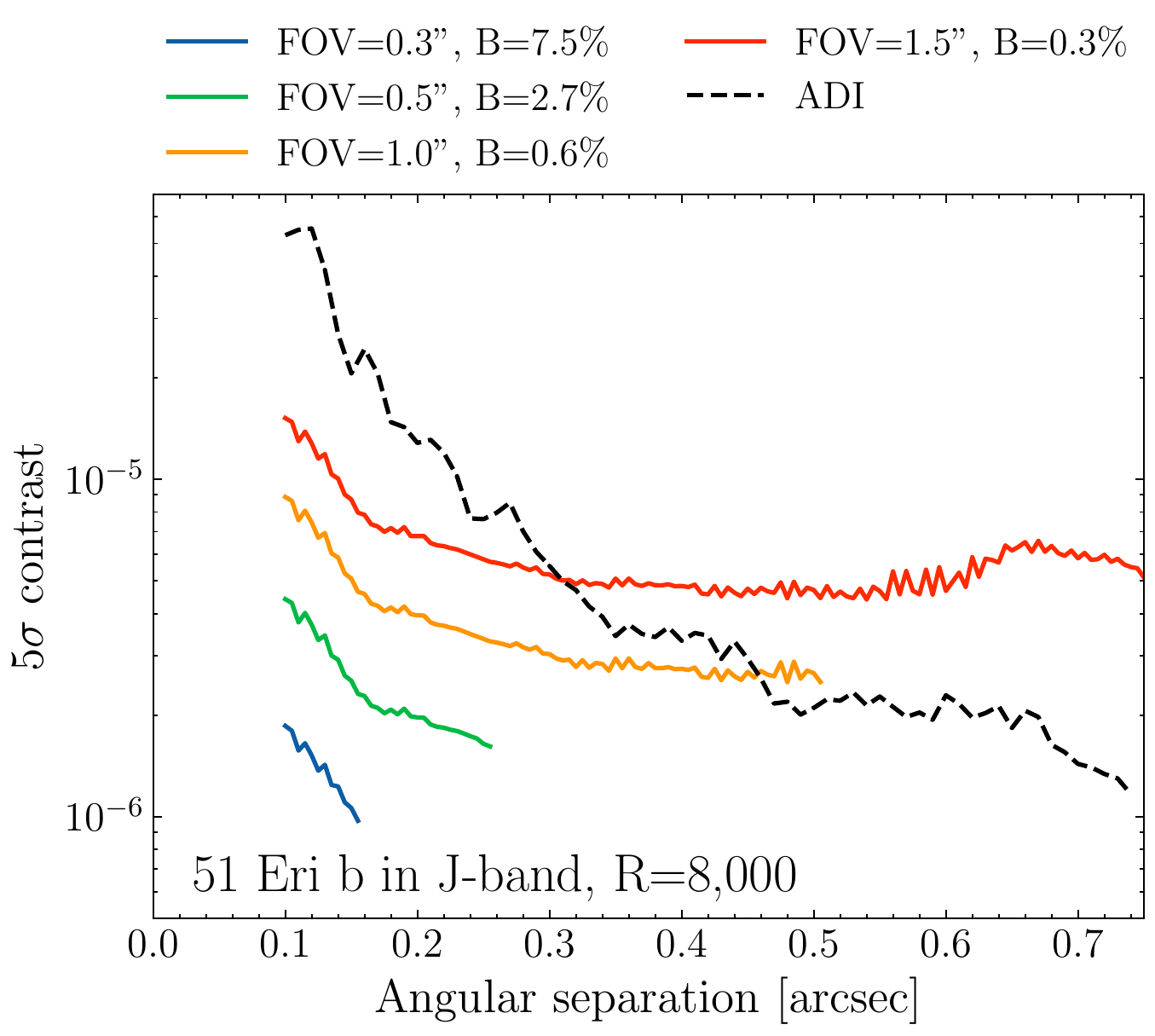}
        \includegraphics[width=0.49\linewidth]{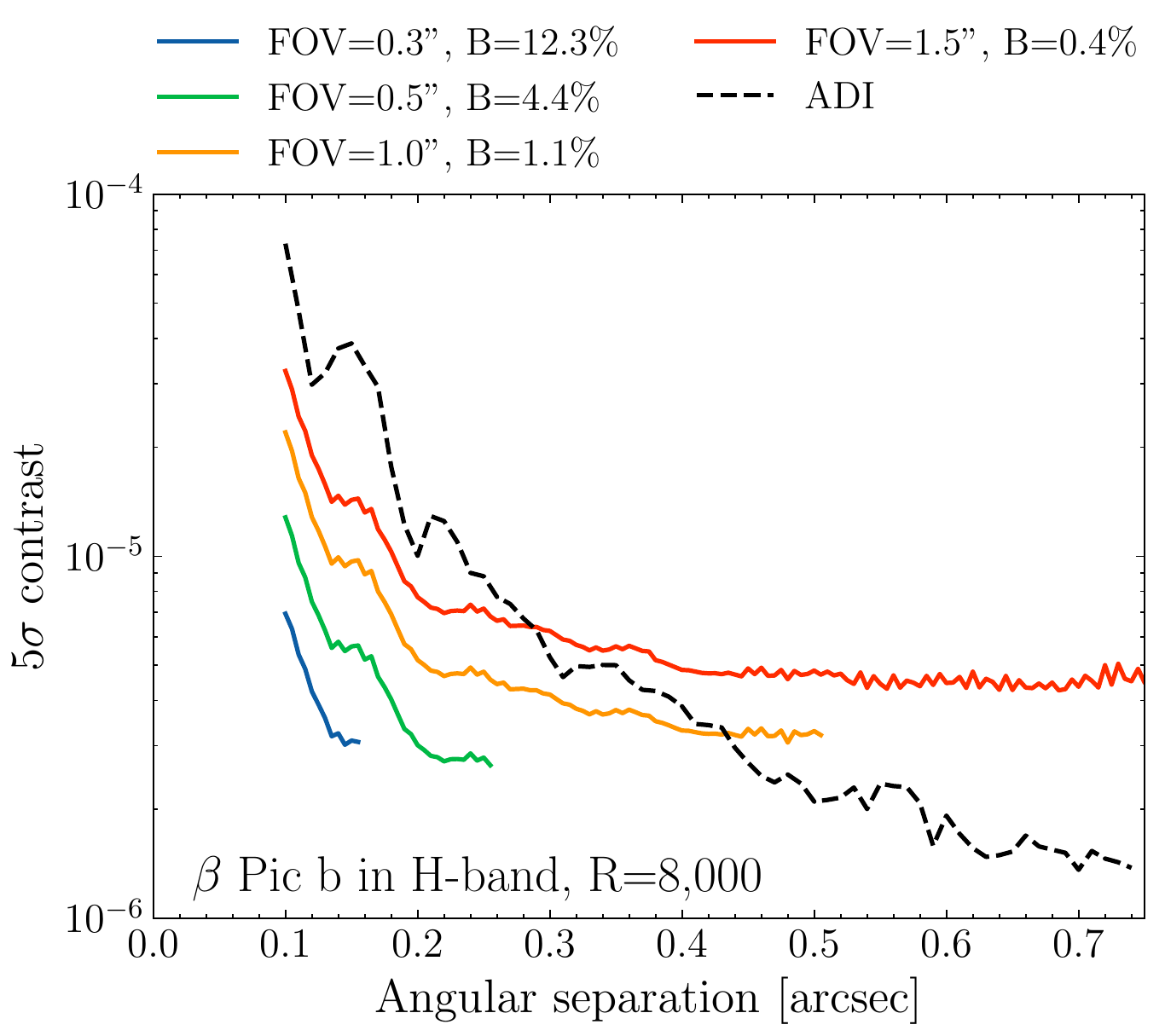}
            \label{fig:contrast_curves}
            \caption{Same as Fig. \ref{fig:contrast_curves} but the spectral resolution is fixed at R=8,000 while varying the spectral bandwidth.}
        \end{figure*}

\section{Validation on VLT/SINFONI data}\label{sec:sinfoni}
The conclusions from the previous sections were based on analytical arguments or idealised end-to-end simulations. However, real observations have more complexity. For example, correlated noise may be present, which may be more prominent at certain spectral resolutions. While there is currently not an IFS with a spectral resolution greater than 1,000 behind a high-contrast imaging system, molecule mapping has been used to successfully detect molecules in $\beta$ Pic b and the HR8799 planets using VLT/SINFONI and Keck/OSIRIS, respectively \citep{2018A&A_hoeijmakers_mm, Ruffio_2021AJ_HR8799_deep}, even though they lack the stellar suppression that can be achieved through XAO and coronagraphy. To validate our found trade-offs, we made use of archival VLT/SINFONI data of $\beta$ Pic b taken on 10 September 2014 originally published in \citet{2018A&A_hoeijmakers_mm}.

\subsection{Data and analysis}\label{sec:sinfoni_data}
    The SINFONI dataset consisted of 24 science frames of the $\beta$ Pictoris system with exposure times of 60 seconds in four dithering positions. Each spaxel in the datacube covered 0.0125'' by 0.025'', and the spectra had a spectral resolution of $\sim 4,500$. The star was placed outside the FOV in these observations to allow for longer exposures. Furthermore, they were taken in pupil tracking mode to facilitate the application of ADI. We followed the data reduction process of \citet{2018A&A_hoeijmakers_mm}. First, a master spectrum was generated from the 20 brightest spaxels. To obtain a model of the modulation of the stellar spectrum at each spaxel, we divided the signal in each spaxel by this master spectrum and subsequently low-pass filtered it by convolving with a Gaussian kernel with a standard deviation of 0.01 $\mu$m. This modulation was then multiplied with the master spectrum again and was subtracted from each spaxel. This method does not consider line spread function variations across the FOV. Furthermore, if the planet is really close to the star, spectral features of the planet may leak into the master spectrum, leading to self-subtraction. To remove correlated high-order structures, we did       a PCA on the spectra of all spaxels and subtracted the first few fitted PCA modes. After this, we used Eq. \ref{eq:mf_snr} to calculate the matched filter S/N for three different high-pass filtered templates: a BT-SETTL model with $T_{\textrm{eff}}$=1700 K and log(g)=4.0 and templates of the contribution of H$_2$O and CO. We again used a Gaussian for the spatial part of the matched filter, with the standard deviation obtained from fitting the stellar PSF. Unlike in the simulations, we did not know the uncertainty of each data point. The uncertainty of each wavelength bin was estimated by taking the standard deviation over all the spaxels. After the matched filter, we empirically normalised the detection map by calculating the mean and standard deviation of all the spaxels within a radius of 15 pixels of the planet while excluding a radius of 7 pixels around the centre of the planet signal. The mean was then subtracted, and we divided by the standard deviation. While this allowed for an estimate of the S/N, the accurate estimation of the true detection significance in real observations remained complex.

\subsection{Spectral resolution}
\begin{figure*}[htbp]
        \centering
        \includegraphics[width=\linewidth]{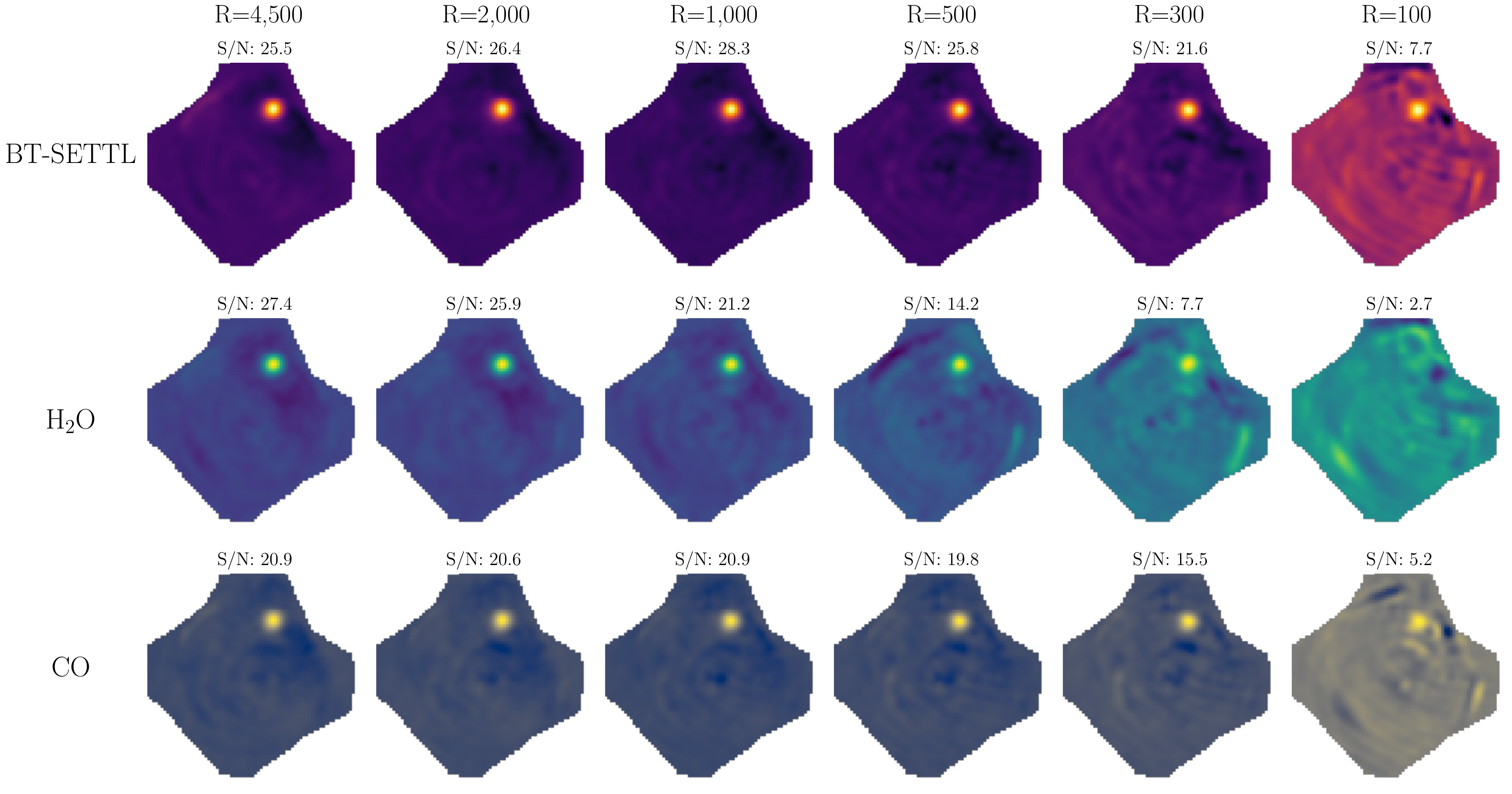}
        \caption{Detection maps of the $\beta$ Pictoris SINFONI data. Each column shows the detection map at a different spectral resolution, while the rows are for different cross-correlation templates.}
        \label{fig:sinfoni_ccf_maps}
    \end{figure*}

    In Section \ref{sec:sim_R}, we studied the effect of the spectral resolution under idealised circumstances, that is, where the stellar and telluric contributions are perfectly removed. However, a higher spectral resolution may improve our ability to remove these contributions. We therefore studied the effect of spectral resolution on real observations. Before applying the data analysis described in Section \ref{sec:sinfoni_data}, we degraded the cubes to different spectral resolutions. This was done by convolving with a Gaussian kernel with an FWHM corresponding to that spectral resolution and then downsampling the spectrum for each spaxel. Molecule mapping was then applied to each of these datasets. Since we had a different number of data points for each spectral resolution, the optimal number of PCA components to apply would potentially also be different. Therefore, we subtracted one, three, five, eight and ten modes and chose the one that gave the highest detection S/N. The resulting detection maps for different spectral resolutions and the three templates are shown in Fig. \ref{fig:sinfoni_ccf_maps}. This figure shows that even if the spectral resolution of the instrument would have been $\sim 300$, we would have had very strong detections of both water and CO.

\begin{figure}[htbp]
        \centering
        \includegraphics[width=0.9\linewidth]{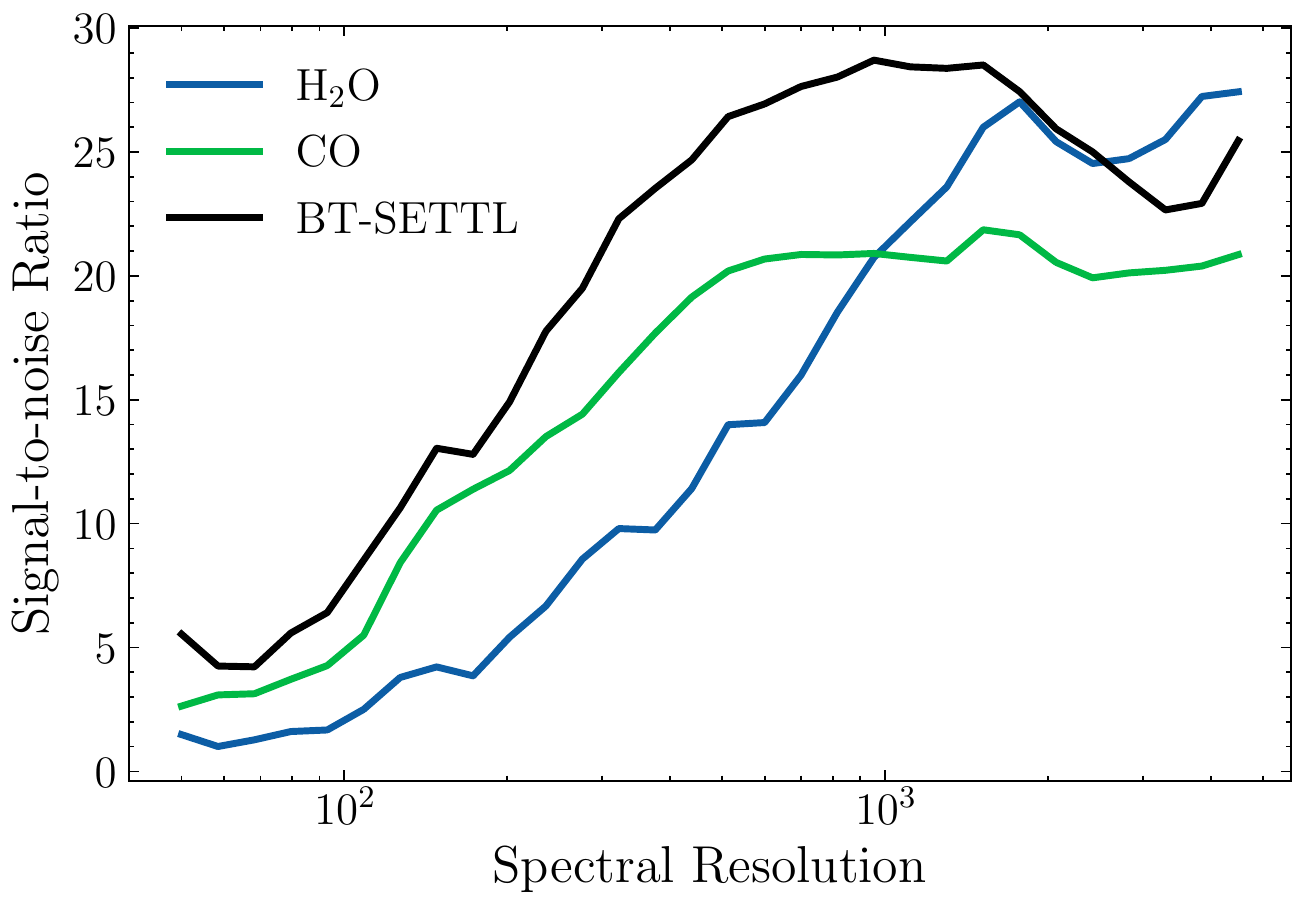}
        \caption{Detection S/N of $\beta$ Pictoris b in the SINFONI data as a function of the spectral resolution to which all datacubes are convolved for H$_2$O, CO, and BT-SETTL templates.}
        \label{fig:sinfoni_snr_vs_R}
    \end{figure}

       Fig. \ref{fig:sinfoni_snr_vs_R} shows the peak detection S/N as a function of spectral resolution. The results are similar to the ones obtained in Section \ref{sec:sim_snr_vs_R}, even though the S/N estimation here may be less accurate. The figure again shows that CO can be detected at lower spectral resolutions than water. We also note that the S/N does not increase after $R\sim 2,000$ and even decreases for the BT-SETTL model. This could be the result of an imperfect wavelength solution for each of the exposures, effectively decreasing the true spectral resolution of the observations. Alternatively, it could be due to inaccuracies in the used line lists, which is more important at higher spectral resolutions.
       
\subsection{Spectral resolution versus bandwidth}
\begin{figure}[htbp]
        \centering
        \includegraphics[width=0.9\linewidth]{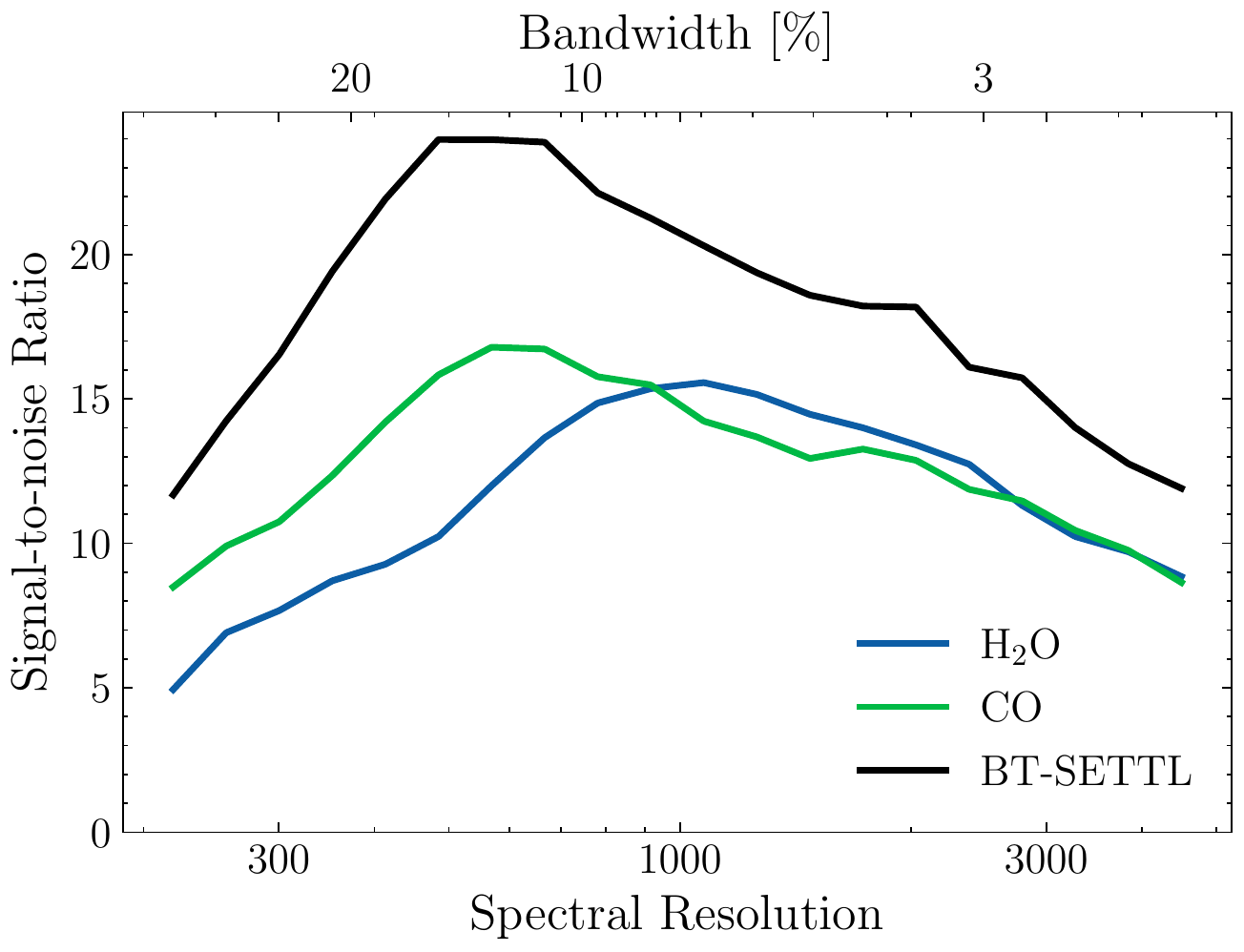}
        \caption{Detection S/N of $\beta$ Pic b in the SINFONI data in the case where the number of spectral bins is fixed. A higher spectral resolution therefore means a lower spectral bandwidth.}
        \label{fig:sinfoni_R_vs_B}
    \end{figure}

    We also validated the results from the spectral resolution versus bandwidth trade-off obtained in Section \ref{sec:sim_R_vs_B}. We again degraded the spectra of all spaxels to different spectral resolutions but adjusted the bandwidth such that the total number of spectral bins remained constant. We used a total of 100 spectral bins per spaxel. We did the S/N calculation for ten different central wavelengths linearly separated between 2.32 and 2.38 $\mu$m and took the mean. This wavelength range ensured that we had contributions from both water and CO. The results of this trade-off are shown in Fig. \ref{fig:sinfoni_R_vs_B}. We observed that the peak S/N for CO is obtained around R$\sim 600$, while for water it is around R$\sim1,000$. The curve for the water signal is in good agreement with what we found in Section \ref{sec:sim_R_vs_B}. For CO, we found a slightly higher optimal spectral resolution. This is the result of the CO features not being distributed over the entire wavelength range such that we did not gain anything from increasing the spectral coverage beyond a certain point. Furthermore, at $R\sim 100$ we may still be in the speckle-limited regime. In this case, the BT-SETTL S/N also peaked at the same spectral resolution as CO. The reason for this is that CO is a more dominant contributor to the spectrum for hotter planets in the K-band than water, as was found in Section \ref{sec:snr_vs_wavelength}. In contrast, the spectrum studied in Section \ref{sec:sim_R_vs_B} was dominated by water features.
    
\section{Atmospheric characterisation}\label{sec:characterisation}
    Next to improving the detection limits of the instrument, an IFS immediately provides a spectrum of the planet which can be used to characterise its atmosphere. A common approach to infer properties about exoplanet atmospheres is through retrievals. The retrievals use a forward model and a sampler to obtain the posterior distributions of the parameters given the observed spectrum. This approach has successfully been applied for low-, medium-, and high-resolution spectra of a directly imaged planet in order to infer its atmospheric properties \citep[e.g.][]{2017A&A_samland_51eri, 2020A&A_molliere_hr8799, 2021Natur_zhang_isotope,Wang_2021AJ_hr8799_kpic}. Assuming a Gaussian likelihood function $\mathcal{L}$, we have:
    \begin{equation}
        \ln{\mathcal{L}} \propto -\sum_i \frac{(s_i-t_i)^2}{\sigma_i^2} \propto \sum_i \frac{s_i t_i}{\sigma_i^2}.
    \end{equation}
    The latter term is the same as in Eq. \ref{eq:mf_snr}, which gives the S/N of a matched filter. The relation between likelihood and cross-correlation is discussed in more detail in \citet{2019AJ_Brogi_retrieval} and \citet{2019AJ_ruffio_radial_velocity}.
    
    When inferring properties of exoplanet atmospheres, the absolute value of the likelihood is not of interest. Instead, the difference in likelihoods between two models is of interest, with parameters $\theta$ and $\theta+\Delta\theta$:

\begin{equation}
    \begin{split}
    \Delta \ln{\mathcal{L}} =& \ln{\mathcal{L}}(\theta)-\ln{\mathcal{L}}(\theta + \Delta \theta) \\
    \propto& \sum_i \frac{s_i}{\sigma_i^2}(t_i(\theta)-t_i(\theta + \Delta \theta)).
 \end{split}
\end{equation}
The obtained constraints on the parameters thus depends on how much the template changes as a function of the targeted parameter. It is not straightforward to derive this analytically as a function of, for example, spectral resolution. Instead, the effect of the spectral resolution on the obtained confidence intervals of different parameters is numerically studied. We setup a retrieval framework using \texttt{petitRADTRANS} \citep{2019A&A_molliere_prt} and \texttt{emcee} \citep{2013PASP_foreman_mackey_emcee}. We used the following free parameters in our forward model: C/O ratio, metallicity, log(g), a flux scaling factor determined by the radius of the planet and distance, the radial velocity of the planet, and its spin along the line of sight vsin(i). Finally, we parameterised the pressure-temperature (P-T) profile of the planet using four free points logarithmically distributed between 0.02 bar and 5 bar. We did not allow for temperature inversions and obtained the full P-T profile using cubic spline interpolation from the four points. We used the mock observations of HR8799 e in the H-band from Section \ref{sec:mock_obs}. Since the data analysis generally results in the loss of the planet continuum, we retrieved the continuum-removed planetary spectrum. We ran a chain of 100 walkers for 3000 steps and subsequently calculated the 68\% confidence interval for the last 1000 steps. The obtained 68\% confidence intervals for the different parameters are shown in Fig. \ref{fig:retrieval_constraints} and an example corner plot of the obtained posteriors is shown in Appendix \ref{app:corner_plot}. The constraint on the temperature profile is the averaged uncertainty interval on the four temperature points. For comparison, we overplotted the confidence intervals with a scaled and fitted version of the inverse matched filter S/N. If the S/N ratio of the planet detection increased, this led to an increase in the maximum likelihood, and we thus expected the confidence interval to shrink. 

We observed that for the C/O ratio, metallicity, surface gravity, and the temperature profile, the dependency on spectral resolution followed that of the matched filter S/N. We therefore argue that trade-offs from the previous sections also hold for the constraints on these parameters. As expected, different behaviour was found for the radial velocity and spin of the planet. For the radial velocity, we saw an $(R \times S/N)^{-1}$ dependence as a result of the smaller resolution element and increase in S/N. The constraint on the spin of the planet was more complex. At a low spectral resolution, this constraint is just an upper limit, as we do not resolve the lines themselves. From $R\sim 8,000$, the line shape starts to be dominated by the rotational broadening, as opposed to the instrumental profile, leading to a sudden decrease in the uncertainty on vsini. An important caveat is that we did not include clouds in these retrievals, which are known to lead to degeneracies at a lower spectral resolution. A more thorough analysis of the model degeneracies and parameter information content \citep{2012ApJ...749...93L_line_informationcontent, 2017AJ....153..151B_batalha_information_content_jwst} at different spectral resolutions and instrument configurations will be the subject of future work.

    \begin{figure}
        \centering
        \includegraphics[width=\linewidth]{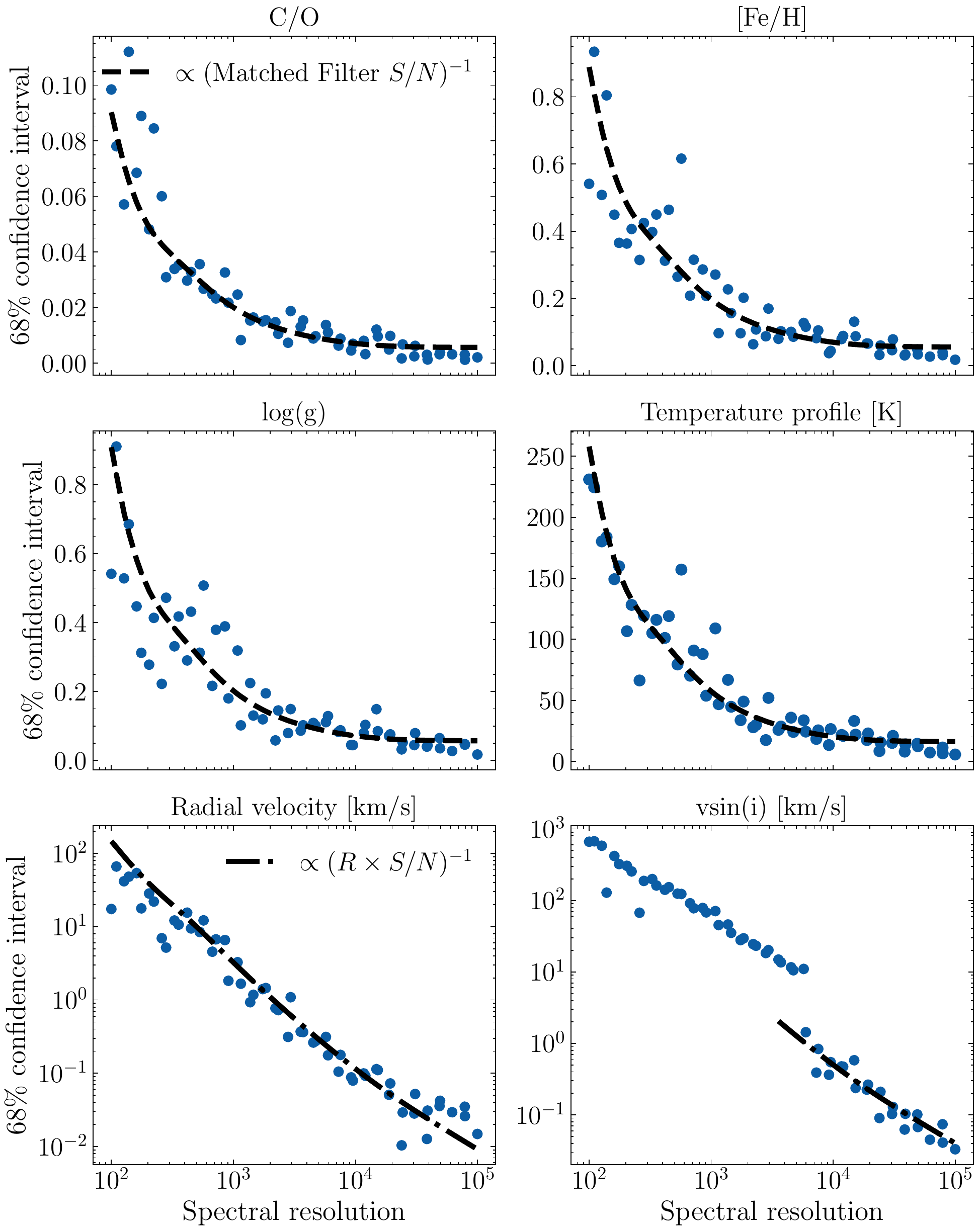}
        \caption{Constraints on the free parameters from the retrievals as a function of spectral resolution. Also shown is the inverse matched filter S/N for the top two rows and a $(R \times \textrm{S/N})^{-1}$ relation for the bottom row.}
        \label{fig:retrieval_constraints}
    \end{figure}
    
\section{Conclusions}
We have studied the trade-offs between the spectral resolution, spectral bandwidth, and FOV for the detection and characterisation capabilities of an IFS behind a high-contrast imaging system. This was studied through end-to-end simulations, analytical considerations, and atmospheric retrievals. The results were then verified on archival data of $\beta$ Pic b with VLT/SINFONI. While we have mainly considered the molecule mapping framework from \citet{2018A&A_hoeijmakers_mm}, our results should be independent of the exact analysis framework (e.g. \citet{2019AJ_ruffio_radial_velocity}), except for the results on the SINFONI data. The main conclusions from this work are the following:

\begin{itemize}
    \item Molecular absorption spectra have decreasing power for higher spectral resolution. Moderate spectral resolutions ($R\gtrsim300$) are therefore already powerful for boosting detection limits with molecule mapping.
    \item In order to get the highest S/N, it is best to increase the spectral resolution until R$\sim$2,000 and then maximise the wavelength coverage of the instrument or observations within the observing band.
    \item Molecule mapping is most beneficial in the speckle-limited regime close to the star. Further away, the loss of the planet continuum results in classical approaches potentially giving similar or better performance. Therefore, such instruments do not need to have a large FOV.
    \item T-type companions are best detectable in the J/H band through water and methane features, while L-type companions are best detectable in the K/H band through CO and water features. The highest detection S/N can often be achieved at the edges of the observing bands.
    \item Constraints on atmospheric parameters such as the C/O ratio, metallicity, surface gravity, and the temperature profile, follow similar trade-offs as detection. Higher spectral resolution is needed to obtain constraints on the radial velocity and spin of the planet.
\end{itemize}

Our results can help guide decisions about instrument designs or observing plans for current and future high-contrast integral field spectrographs in the near-infrared, such as VLT/SPHERE+/MedRes, ELT/HARMONI, or ELT/PCS. Future work will investigate how these trade-offs change when searching for planets in reflected light.

\begin{acknowledgements}
We thank Arthur Vigan for providing us the representative XAO residuals for SPHERE. We also want to thank the referee for suggestions that have resulted in major improvements in this work. R.L. and I.S. acknowledge funding from the European Research Council (ERC) under the European Union's Horizon 2020 research and innovation program under grant agreement No 694513. This work has been supported by a grant from Labex OSUG (Investissements d’avenir – ANR10 LABX56). C.D. is part of Labex OSUG (ANR10 LABX56), and C.D. acknowledges support from the European Research Council under the European Union’s Horizon 2020 research and innovation program under grant agreement No. 832428-Origins.

\end{acknowledgements}

\bibliographystyle{aa}
\bibliography{main}

\begin{appendix}

\section{Spectral models and templates}\label{app:models}
\begin{figure*}[htbp]
    \centering
    \includegraphics[width=\linewidth]{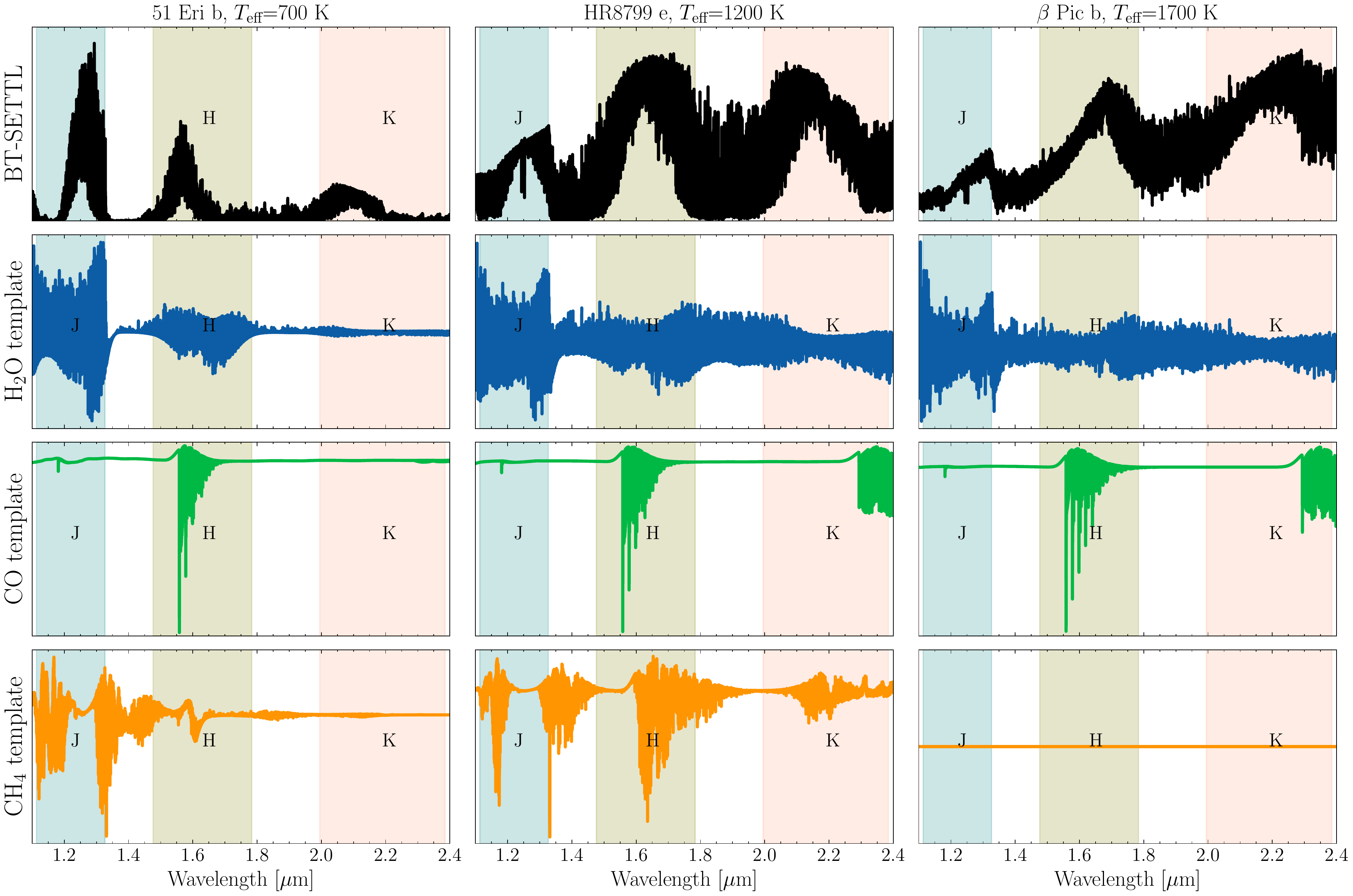}
    \caption{Flux density of the spectral models and templates used throughout this study shown at a spectral resolution of 30,000. The molecular templates have been high-pass filtered.}
    \label{fig:spectral_models}
\end{figure*}

The spectral models and templates used throughout this work are shown in Fig. \ref{fig:spectral_models} for a spectral resolution of 30,000.

\section{Power spectral densities for different planets}\label{app:psd}
\begin{figure}[htbp]
    \centering
    \includegraphics[width=\linewidth]{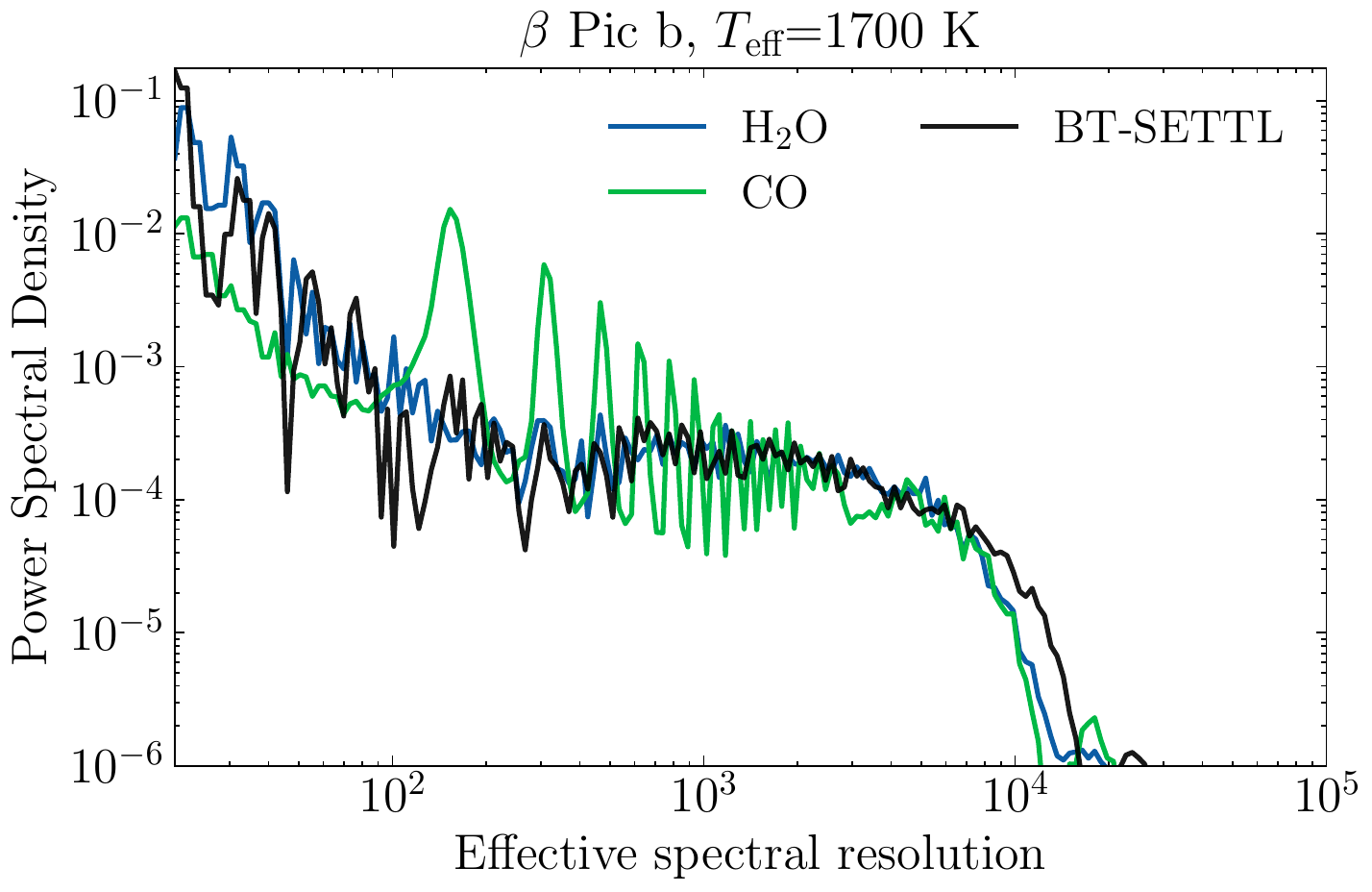}
    \includegraphics[width=\linewidth]{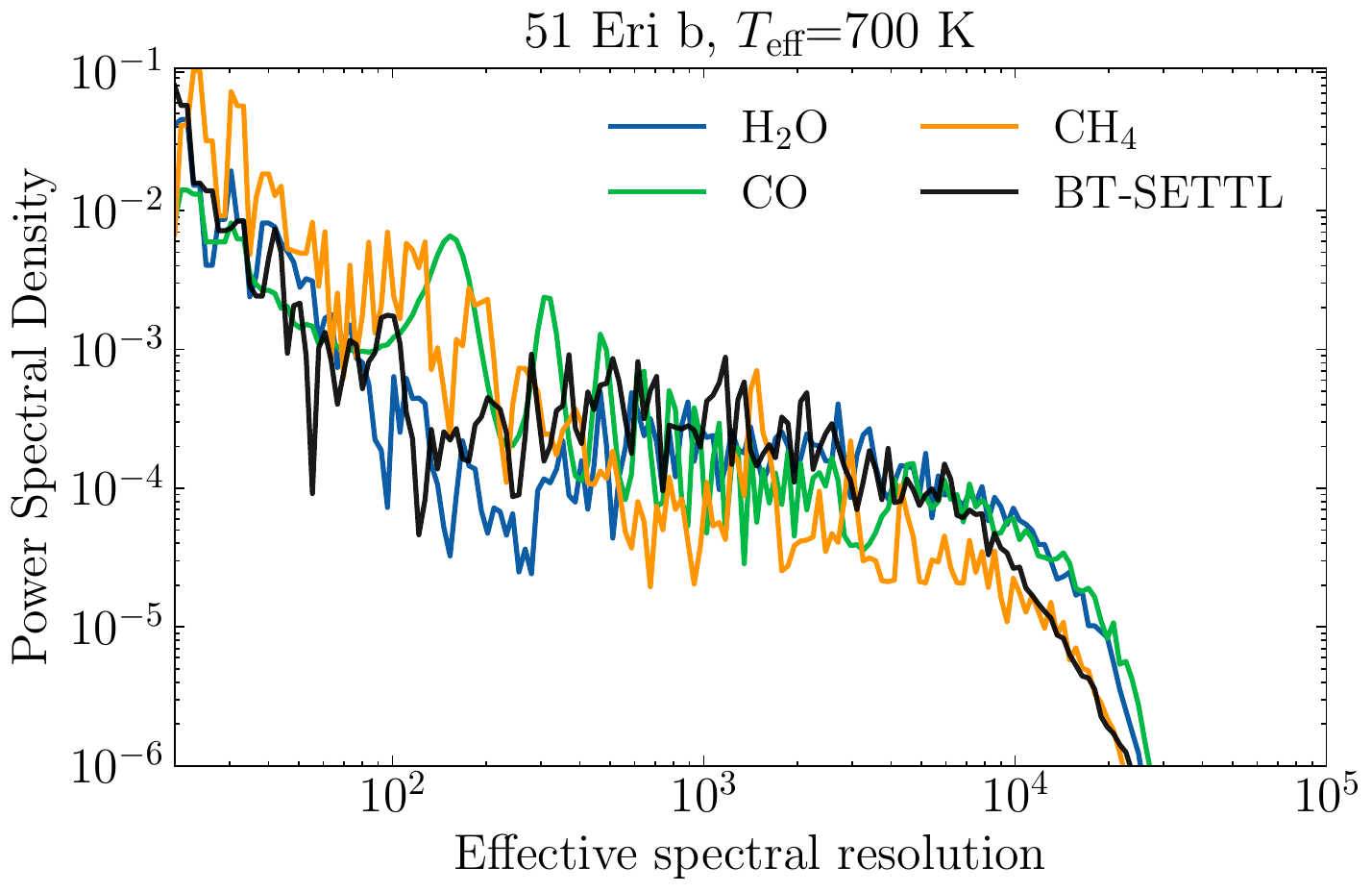}
    \caption{Power spectral densities of the planetary and molecular templates for a $\beta$ Pic b-like (top) and 51 Eri b-like (bottom) planets.}
    \label{fig:PSD_diff_planets}
\end{figure}

Fig \ref{fig:PSD_diff_planets} shows the PSDs of $\beta$ Pic b-like and 51 Eri b-like planets.

\section{Derivation of the relation between the cross-correlation S/N and the power spectral density}\label{app:derivations}
Given a spectrum $s_i$, template $t_i$, and Gaussian noise with standard deviation $\sigma_i$, the S/N of a matched filter is given by \citep{2017ApJ_ruffio_fmmf}:
\begin{equation}
    S/N = \frac{\sum_i^N s_i t_i/\sigma_i^2}{\sqrt{\sum_i^N t_i^2/\sigma_i^2}},
\end{equation}
where $N$ is the number of data points indexed by $i$. For simplicity in these derivations, we assumed that the planet signal is concentrated in a single spaxel with $M$ spectral bins. Assuming we have a perfect model ($s_i = t_i + n$) and uncorrelated noise that is roughly constant for each datapoint ($\sigma_i \approx \sigma$), we have:
\begin{equation}\label{eq:B2}
    S/N \approx \frac{1}{\sigma} \sqrt{\sum_i^M t_i^2}.
\end{equation}
In the case that $t_i$ is sampled at a sufficient rate such that we can reconstruct the continuous signal $t$, (i.e. we Nyquist sample the spectrum), we have $\sum t_i^2 = M <t^2>$, with $<t^2>$ denoting the average value of $t^2$. This gives:
\begin{equation}\label{eq:mf_snr2}
     S/N \approx \frac{1}{\sigma} \sqrt{M<t^2>}.
\end{equation}
We defined the wavelength sampling such that we have constant spectral resolution across the entire wavelength range (i.e. $(\lambda_{i+1}-\lambda_i)/\lambda_i = \textrm{constant}$). We did this by defining a wavelength grid uniformly in $\log\lambda$. We can use the Wiener-Khinchin theorem to express this in terms of the PSD:
\begin{equation}\label{eq:Khinchin}
    <t^2(\log \lambda) > = R_t(0)= \int_{-\infty}^\infty \textrm{PSD}[t(\log \lambda)](f) df.
\end{equation}
Here, $R_t$ is the autocorrelation function of $t(\log \lambda)$, $\textrm{PSD}[t(\log \lambda)]$ is the PSD of $t(\log \lambda)$ as defined in Eq. \ref{eq:planet_psd_def}, and $f=1/(\Delta \log \lambda')$ is the frequency of the corresponding fluctuation. One can also express the PSD in terms of an effective spectral resolution $R'$ by using:
\begin{equation}
\begin{split}
     \Delta \log\lambda' &= \log(\lambda + \Delta \lambda') - \log(\lambda) \\
     &= \log \left( \frac{\lambda + \Delta \lambda'}{\lambda} \right) \\
     &=\log \left( 1+ \frac{\Delta \lambda'}{\lambda} \right) \\
     & \approx \Delta \lambda'/\lambda = 1/(2R') \quad \textrm{for} \hspace{5pt} \Delta \lambda'/\lambda \ll 1.
\end{split}
\end{equation}
The factor of two comes from the fact that we needed to Nyquist sample the spectrum to see fluctuations of a specific frequency. This means that we have $R' = f/2$, $df = 2dR'$, and that $\log \lambda$ and $2R'$ are Fourier conjugate variables. Furthermore, we can use that $t$ is real-valued to turn Eq. \ref{eq:Khinchin} into a one-sided integral with an additional factor of two. Together this gives:
\begin{equation}
    <t^2>=4\int_{0}^{\infty} \textrm{PSD}[t](R') dR'.
\end{equation}

The noiseless observed spectrum $t$ consists of the planet spectrum $t_p$ convolved with the line spread function $\textrm{lsf}_R$ of the instrument, the detector sampling function $s_R$, and a filter describing the effect of the data cleaning $c$ (e.g. the high-pass filter):
\begin{equation}
    t = t_p \ast \textrm{lsf}_R \ast s_R \ast c.
\end{equation}
These convolutions become multiplications in Fourier space:
\begin{equation}
    T(R') = T_p(R') \cdot \textrm{LSF}_R(R') \cdot S_R(R') \cdot C(R'),
\end{equation}
or equivalently:
\begin{equation}
    T(R') = H(R') T_p(R'),
\end{equation}
where $H(R')= \textrm{LSF}_R(R') S_R(R') C(R')$ is the combined transfer function of the spectrograph and data reduction. We then have:
\begin{equation}
    <t^2> = 4 \int_{0}^{\infty} |H(R')|^2 P(R') dR',
\end{equation}
where $P(R')= \textrm{PSD}[t_p]$ is the PSD of the planet spectrum as defined in Eq. \ref{eq:planet_psd_def}. Substituting this into Eq. \ref{eq:mf_snr2}, we obtain:
\begin{equation}
    S/N \approx \frac{2\sqrt{M}}{\sigma}\sqrt{\int_0^{\infty}|H(R')|^2 P(R') dR'}.
\end{equation}
In the Nyquist-sampled case, we have $M=2 B R$, with $B=(\lambda_{\textrm{max}} - \lambda_{\textrm{min}})/\lambda_{\textrm{central}}$ the spectral bandwidth, giving:
\begin{equation}\label{eq:snr_with_filter_app}
    S/N \approx \frac{2\sqrt{2BR}}{\sigma}\sqrt{\int_0^{\infty}|H(R')|^2 P(R') dR'}.
\end{equation}

Next, we describe some simplifying assumptions we made on the filter $H(R')$. We first assumed that we have an idealised spectrograph where the line spread and sampling function together as an ideal low-pass filter, removing all fluctuations with frequencies higher than R:
\begin{equation}
    \textrm{LSF}_R (R') S(R') =
\begin{cases}
    1/R& \text{if } R'\leq R\\
    0              & \text{otherwise}.
\end{cases}
\end{equation}
Here, the $1/R$ is to make sure that the line spread function is normalised and the flux is conserved. Finally, we assumed that the data reduction is an ideal high-pass filter that removes all features below $R'=R_{min}$. This then simplifies \ref{eq:snr_with_filter_app} to:
\begin{equation}
    S/N \approx \frac{2\sqrt{2B}}{\sigma\sqrt{R}}\sqrt{\int_{R_{\textrm{min}}}^{R} P(R') dR'}.
\end{equation}


\section{Numerical validation of the matched filter S/N equation}\label{app:equation_validation}
\begin{figure*}
    \centering
    \includegraphics[width=0.95\linewidth]{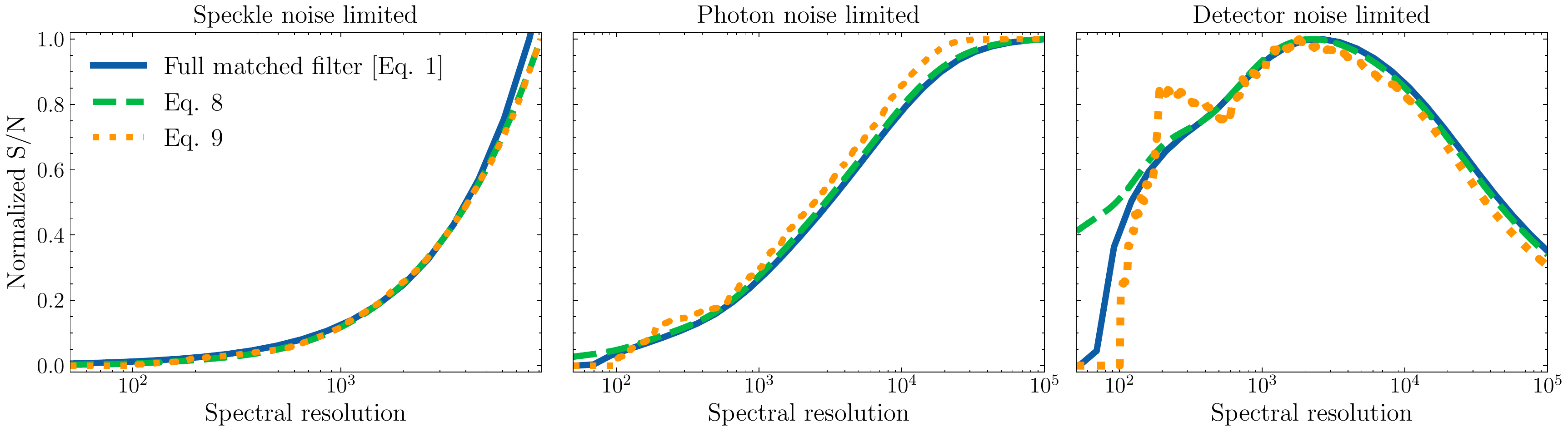}
    \caption{Numerical validation of the relation between the PSD and the S/N of the matched filter.}
    \label{fig:snr_validation}
\end{figure*}

Here we show a numerical validation of the equation derived in Appendix \ref{app:derivations}. We used the mock observations of HR8799e in the H-band as described in Section \ref{sec:mock_obs} as a test case. We calculated the cross-correlation S/N in three ways: 1) The first method was by applying Eq. \ref{eq:mf_snr} on the full simulated datacube. This is the baseline and most accurate case, as it includes the most effects. In the speckle-noise limited case this equation is not applicable, and we calculated the S/N by normalizing by the standard deviation of the cross-correlation function away from the peak (as in e.g. \citet{2018A&A_hoeijmakers_mm, 2014Natur_snellen_beta_pic_crires}). 2) The second approach used Eq. \ref{eq:snr_psd} based on the PSD, which assumes wavelength independent errors. For this method, we used Gaussians for $\textrm{LSF}(R')$ and $C(R')$ with appropriate standard deviations, given the spectral resolution and high-rpass filter applied to the data. 3) The third method used Eq. \ref{eq:snr_psd_ideal_lsf}, which makes further simplifying assumptions on the properties of the spectrograph and data reduction. The resulting S/N curves of the methods are shown in Fig. \ref{fig:snr_validation}. It shows good agreement between the three equations, except at very low and very high spectral resolutions, where some of the assumptions break down.

\section{Example retrieval result}\label{app:corner_plot}
\begin{figure*}[htbp]
    \centering
    \includegraphics[width=0.9\linewidth]{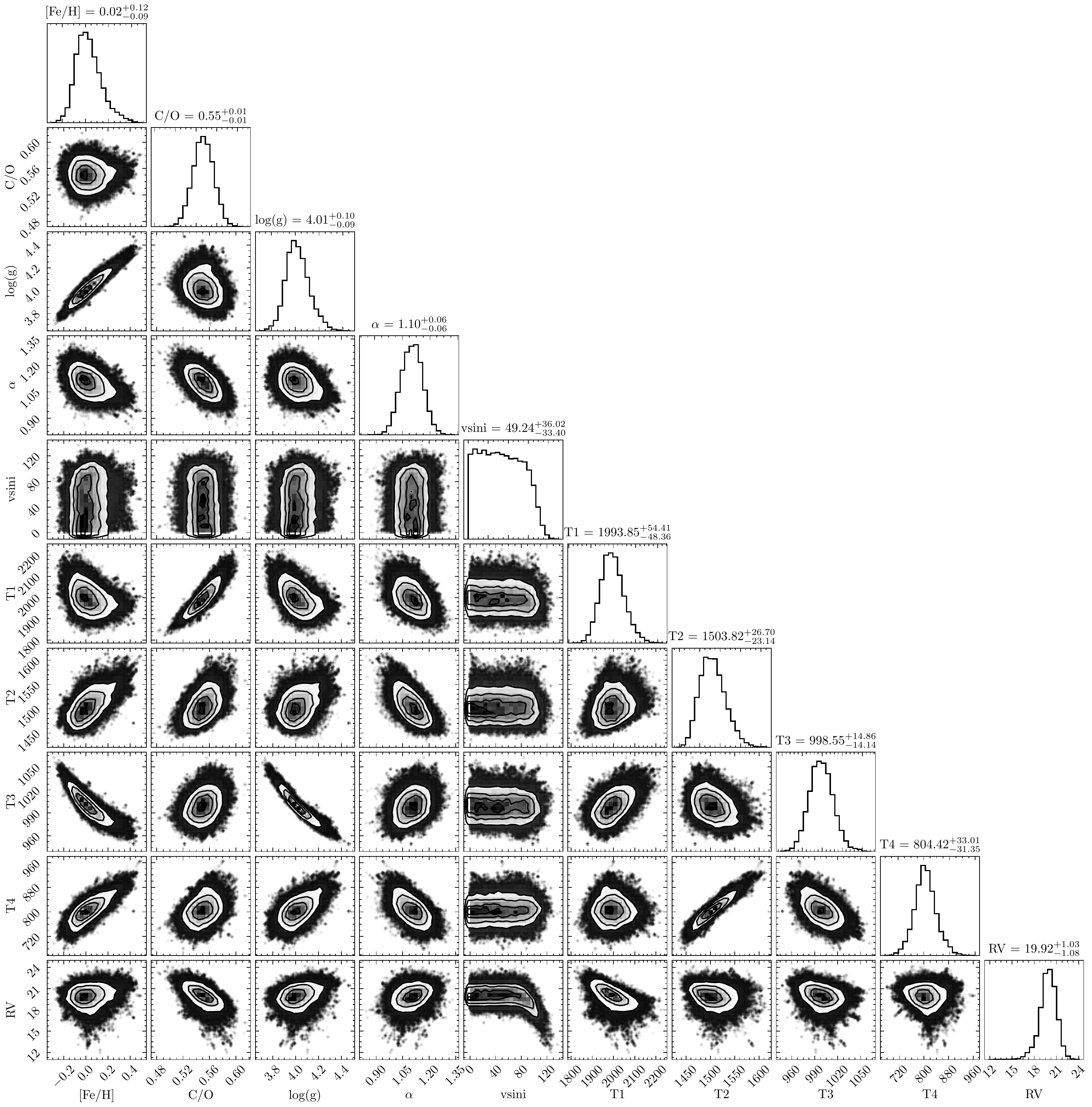}
    \caption{Example corner plot for a retrieval of HR8799e in the H-band with a spectral resolution of 1,000.}
    \label{fig:corner_example}
\end{figure*}

An example corner plot from the retrieval study for HR8799e with a spectral resolution of 1,000 in the H-band is shown in Fig. \ref{fig:corner_example}. The free parameters are the metallicity [Fe/H], C/O ratio, surface gravity log(g), scaling parameter $\alpha$, rotational velocity vsini, temperature nods $T_1$ through $T_4$, and the radial velocity (RV). 
\end{appendix}

\end{document}